\documentclass[apj,twocolumn,twocolappendix,numberedappendix]{openjournal}
\usepackage{amsmath}
\usepackage{booktabs}
\usepackage{multirow}
\usepackage{color}
\usepackage{soul}
\usepackage{booktabs} 
\usepackage{threeparttable}

\usepackage{siunitx}

\usepackage[dvipsnames]{xcolor} 

\usepackage[breaklinks,colorlinks,citecolor=blue,urlcolor=blue]{hyperref}
\usepackage{xurl}

\defcitealias{Desai24}{D24}

\renewcommand{\arraystretch}{1.2}
\usepackage{listings}
\usepackage{color}
\definecolor{dkgreen}{rgb}{0,0.6,0}
\definecolor{gray}{rgb}{0.5,0.5,0.5}
\definecolor{mauve}{rgb}{0.58,0,0.82}
\definecolor{golden}{rgb}{0.86,0.65,0.01}
\usepackage{orcidlink}

\begin{document}


\title{\textbf{S\lowercase{upernova} R\lowercase{ates and} L\lowercase{uminosity} F\lowercase{unctions from} ASAS-SN III: \\O\lowercase{ver a} D\lowercase{ecade of} T\lowercase{ype} I\lowercase{a} SN\lowercase{e and} T\lowercase{heir} S\lowercase{ubtypes}}}

\author{\vspace{-1.4cm}
        D.~D.~Desai\,\orcidlink{0000-0002-2164-859X}$^{1}$}
\author{B.~J.~Shappee\,\orcidlink{0000-0003-4631-1149}$^{1}$}
\author{C.~S.~Kochanek\,\orcidlink{}$^{2,3}$}
\author{K.~Z.~Stanek$^{2,3}$}
\author{C.~Ashall\,\orcidlink{0000-0002-5221-7557}$^1$}
\author{J.~F.~Beacom\,\orcidlink{0000-0002-0005-2631}$^{4,2,3}$}
\author{C.~R.~Burns\,\orcidlink{0000-0003-4625-6629}$^5$}
\author{A.~Do\,\orcidlink{0000-0003-3429-7845}$^6$}
\author{Subo~Dong\,\orcidlink{0000-0002-1027-0990}$^{7,8,9}$}
\author{W.~B.~Hoogendam\,\orcidlink{0000-0003-3953-9532}$^{1}$}
\author{J.~Lu\,\orcidlink{0000-0002-3900-1452}$^{10}$}
\author{T.~Pessi\,\orcidlink{0000-0001-6540-0767}$^{11}$}
\author{J.~L.~Prieto$^{12,13}$}
\author{T.~A.~Thompson\,\orcidlink{0000-0003-2377-9574}$^{2,4,3}$\vspace{0.15cm}}

\affiliation{$^1$ Institute for Astronomy, University of Hawai‘i, 2680 Woodlawn Drive, Honolulu, HI 96822, USA}

\affiliation{$^{2}$ Department of Astronomy, The Ohio State University, 140 West 18th Avenue, Columbus, OH 43210, USA}

\affiliation{$^{3}$ Center for Cosmology and AstroParticle Physics, The Ohio State University, 191 West Woodruff Avenue, Columbus, OH 43210, USA}

\affiliation{$^{4}$ Department of Physics, The Ohio State University, 191 West Woodruff Avenue, Columbus, OH 43210, USA}

\affiliation{$^{5}$ Observatories of the Carnegie Institution for Science, 813 Santa Barbara Street, Pasadena, CA 91101, USA}

\affiliation{$^{6}$ Institute of Astronomy and Kavli Institute for Cosmology, Madingley Road, Cambridge, CB3 0HA, UK}

\affiliation{$^{7}$ Department of Astronomy, School of Physics, Peking University, 5 Yiheyuan Road, Haidian District, Beijing 100871, People's Republic of China}

\affiliation{$^{8}$ Kavli Institute of Astronomy and Astrophysics, Peking University, 5 Yiheyuan Road, Haidian District, Beijing 100871, People's Republic of China}

\affiliation{$^{9}$ National Astronomical Observatories, Chinese Academy of Science, 20A Datun Road, Chaoyang District, Beijing 100101, China}

\affiliation{$^{10}$ Department of Physics \& Astronomy, Michigan State University, East Lansing, MI 48824, USA}

\affiliation{$^{11}$ European Southern Observatory, Alonso de Córdova 3107, Vitacura, Casilla 19001, Santiago, Chile}

\affiliation{$^{12}$ Instituto de Estudios Astrof\'isicos, Facultad de Ingenier\'ia y Ciencias, Universidad Diego Portales, Av. Ej\'ercito Libertador 441, Santiago, Chile}

\affiliation{$^{13}$ Millennium Institute of Astrophysics MAS, Nuncio Monsenor Sotero Sanz 100, Off. 104, Providencia, Santiago, Chile}

\email{Corresponding author: dddesai@hawaii.edu}

\begin{abstract}
We present volumetric rates and luminosity functions (LFs) of Type Ia supernovae (SNe Ia) from the All-Sky Automated Survey for Supernovae (ASAS-SN), covering the 11-year period from 2014 to 2024. By combining the 2014--2017 $V$-band sample with the 2018--2024 $g$-band sample, we construct a large statistical dataset of $1776$ SNe~Ia. We compute completeness corrections based on injection-recovery simulations of the ASAS-SN light curves, taking into account the variations in light curve shapes. For our standard sample ($M_{g,\mathrm{peak}}<-16.0$~mag), we extract a total volumetric SN~Ia rate of $R_{\mathrm{tot}} = (2.55 \pm 0.12) \times 10^4\,\mathrm{yr}^{-1}\,\mathrm{Gpc}^{-3}\,h_{70}^3$ at a median redshift of $z=0.029$. With a statistical uncertainty of $4.7\%$, this is the most precise local measurement to date. While the ``normal'' SNe~Ia account for $(92.7 \pm 1.9)\%$ of this rate, the total LF reveals immense diversity, with $M_{g,\mathrm{peak}}$ spanning over five magnitudes. The LF of SNe~Iax is also broad and rises toward lower luminosities, resulting in a likely lower limit of $(4.3 \pm 1.8)\%$ of the total rate. We place strong constraints on the rate of SNe~Ia-CSM, finding they account for only $(0.036 \pm 0.017)\%$ of the total local rate. Finally, we find that the low-luminosity 02es-like SNe are $7 \pm 5$ times more common than the luminous 03fg-like SNe. This places demographic constraints on models proposing a physical continuum for these two subtypes, implying that any common channel for the two classes must strongly favor lower-luminosity explosions.


\end{abstract}

\maketitle

\section{Introduction} \label{sec:intro}
The rates of Type Ia supernovae (SNe~Ia) and their subtypes are fundamental observables that link stellar evolution to galactic chemical enrichment and cosmology. Precise measurement of these rates constrains progenitor population models, traces the cosmic history of star formation, and quantifies the production of iron-peak elements and Galactic positrons \citep[e.g.,][]{Raiteri96,MatteucciRecchi01,Horiuchi10,maoz-mannucci12,Desai25}. Historically, rate measurements relied on galaxy-targeted searches or heterogeneous compilations, which suffer from small sample sizes and complex selection biases \citep[e.g.,][]{cappellaro99, li11b}. Modern, wide-field, untargeted surveys have transformed this landscape and constructed statistical samples, providing more robust measurements of the local SN~Ia rate \citep[e.g.,][]{frohmaier19, perley20, Srivastav22, sharon-kushnir22, Desai24}.

The All-Sky Automated Survey for Supernovae \citep[ASAS-SN;][]{shappee14, kochanek17, Hart23} provides a uniquely powerful data set. Since 2014, ASAS-SN has observed the entire visible sky with a daily-to-few-day cadence. It remained the only survey with this capability until 2022, when ATLAS expanded its coverage to the Southern hemisphere. ASAS-SN's untargeted strategy and shallow depth ($g \lesssim 18.5$~mag) naturally produces a sample of bright, nearby transients with high spectroscopic completeness, as nearly every discovered object is bright enough for classification with modest telescope resources \citep{Chen22,Tucker22a}. During the 2014--2017 $V$-band phase, ASAS-SN was nearly spectroscopically complete \citep{Holoien17a, Holoien17b, Holoien17c, Holoien19}. While the subsequent $g$-band survey (2018--2024) has a slightly lower average spectroscopic completeness ($\sim81\%$, in part due to issues related to the COVID-19 pandemic), the deeper limiting magnitude and longer time baseline increase the sample by a factor of 3.4 over the $V$-band sample \citep[e.g.,][]{Neumann23}.

This paper is the third in a series from the ASAS-SN team establishing benchmark local transient rates. In Paper I \citep[hereafter \citetalias{Desai24}]{Desai24}, we established our methodology using the 2014--2017 $V$-band SN~Ia sample and found a total volumetric rate of $R_{\mathrm{tot},V} = (2.28 \pm 0.20) \times 10^4\,\mathrm{yr}^{-1}\,\mathrm{Gpc}^{-3}\,h_{70}^3$, a measurement that is statistically precise and systematically robust due to the high completeness of the $V$-band sample. Paper II applied this framework to the core-collapse SNe from the same era \citep{Pessi25}.

Here, we analyze the larger and deeper SN~Ia sample from the ASAS-SN $g$-band survey, covering the years 2018 through 2024. By combining this new sample with the $V$-band sample, we more than quadruple the total sample size. This extended dataset provides the statistical power necessary to dissect the diversity of the SN~Ia population. We provide robust volumetric rates and luminosity functions (LFs) for the major SN~Ia subtypes, including the 91T-like \citep[e.g.,][]{Filippenko92a, Phillips92,Phillips24}, 02es-like \citep[e.g.,][]{Ganeshalingam12}, 03fg-like \citep[e.g.,][]{Howell06,Ashall21}, Iax \citep[e.g.,][]{Foley13, Jha17}, and Ia-CSM events \citep[e.g.,][]{Silverman13,Sharma23}. Quantifying the relative rates of these subtypes is essential for understanding the overall SN~Ia delay-time distribution and distinguishing between progenitor channels \citep[e.g.,][]{Maoz10,Maoz14,White15, Hoogendam24,RuiterSeitenzahl25}.

In Section~\ref{sec:sample}, we describe the $g$-band SN~Ia sample and our light-curve fitting procedure. In Section~\ref{sec:rates}, we detail the methodology for our rate calculations and combining the $V$- and $g$-band samples. We present our results on the volumetric rates in Section~\ref{sec:vol_rates} and the luminosity functions in Section~\ref{sec:LF}. Finally, in Section~\ref{sec:summary}, we summarize our conclusions and their implications. Throughout this analysis, we adopt a flat $\Lambda$CDM cosmology with $H_0 = 70~\mathrm{km~s^{-1}~Mpc^{-1}}$ and $\Omega_{m,0} = 0.3$.

\section{The Supernova Sample} \label{sec:sample}
The SN sample in this paper builds upon the ASAS-SN catalogues presented in \citet{Holoien17a, Holoien17b, Holoien17c, Holoien19} and \citet{Neumann23}. Our added sample covers SNe discovered or recovered by ASAS-SN in the $g$-band between UTC 2021-01-01 and UTC 2024-12-31. To ensure uniformity, the 2018--2020 sample from \citet{Neumann23} is reanalyzed here to provide updated light-curve fits and peak magnitudes consistent with the methods used in \citetalias{Desai24} and \citet{Pessi25}.

Since our time period extends beyond that of \citet{Neumann23}, we compiled a list of SNe discovered by ASAS-SN from the Transient Name Server\footnote{\url{https://www.wis-tns.org/}} (TNS) and the archival ASAS-SN supernova page\footnote{\url{https://www.astronomy.ohio-state.edu/asassn/sn_list.html}}. To identify SNe discovered by other surveys but independently recovered in ASAS-SN data, we cross-matched all transients reported to TNS with the ASAS-SN transient archive\footnote{\url{https://www.astronomy.ohio-state.edu/asassn/transients.html}}. This combined sample of discoveries and recoveries was then filtered to remove non-supernova transients, such as cataclysmic variables, novae, active galactic nuclei, and tidal disruption events. This process resulted in a list of 2703 spectroscopically classified SNe and 1685 unclassified transients. For the unclassified transients, we retained candidates that had a potential host galaxy within a $1\arcmin$ radius and a light curve morphologically consistent with our supernova templates. We further vetted these candidates by cross-matching their positions with stellar catalogues from Gaia, Pan-STARRS, TESS, and GALEX, removing any that had a Galactic stellar counterpart within a $5\arcsec$ radius, thus leaving 652 unclassified potential SNe in our sample.

\begin{figure}
    \centering
    \includegraphics[width=0.9\linewidth]{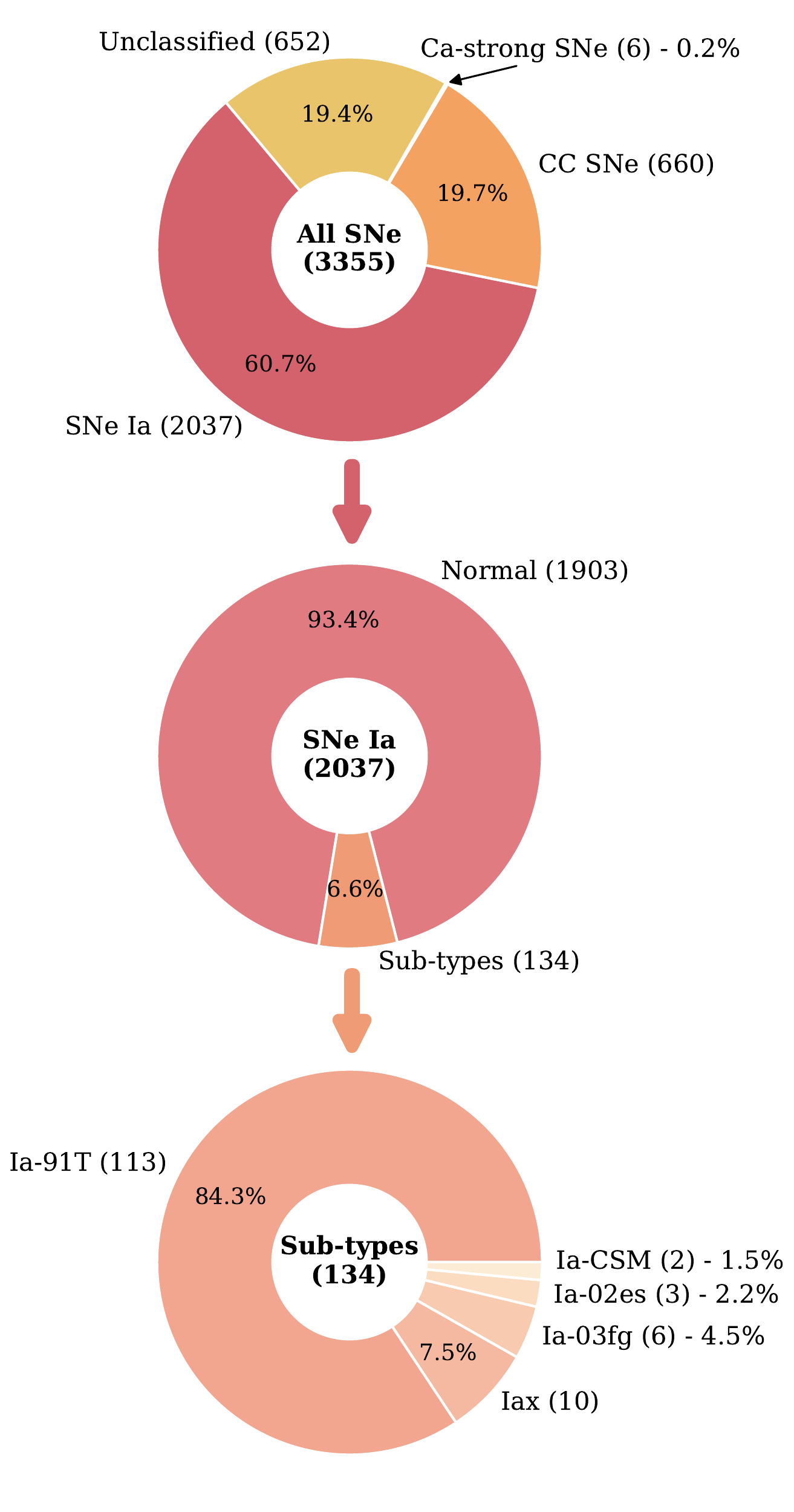}
    \caption{Fractions and numbers of all SNe discovered or recovered by ASAS-SN in $g$-band from UTC 2018-01-01 to UTC 2024-12-31.}
    \label{fig:piecharts}
\end{figure}

We update the classifications for several objects in our sample. ASASSN-18ro (SN~2018evt), originally classified as SN~Ia, is updated to SN~Ia-CSM \citep{Dong18,Yang23}, and SN~2022esa, originally classified as SN~Ia-CSM, is updated to SN~Ic-CSM \citep{Griffith25,Maeda25}. For 11 SNe in our sample classified as SN~Ia-pec on TNS, we examine the spectra to determine more appropriate subtypes: SNe~2019ein, 2020dju, 2021ebb, 2022bbt, and 2022aana are normal SNe~Ia; SNe~2014dt, 2019omz, and 2020rea are SNe~Iax; SNe~2018las, 2021qce, and 2022obk are SNe~Ia-91T-like. We remove AT~2020adgm (ASASSN-20qc) from the sample for being an ambiguous nuclear transient \citep{Pasham24}. Our $g$-band sample comprises 3355 supernovae discovered or recovered by ASAS-SN in the years 2018 to 2024, of which $80.6\%$ are spectroscopically classified. Figure~\ref{fig:piecharts} shows the division of these 3355 SNe into SN types, with 2037 spectroscopically classified SNe~Ia.

We note that distinguishing 91T-like events from the closely related 1999aa-like events and other Branch shallow-silicon supernovae is difficult without a spectral time series \citep{Phillips22}. Consequently, our 91T-like sample likely includes a fraction of these related subtypes. Furthermore, we define the ``normal'' SN~Ia population to encompass both the Core Normal (CN) and Broad Line (BL) Branch subtypes. We also categorize the subluminous 91bg-like subclass and transitional events from TNS as normal SNe~Ia, as recent analyses of these events and the use of the color-stretch parameter $s_{BV}$ demonstrate that they transition smoothly from the luminous population rather than forming a distinct, peculiar class \citep{Ashall18,Burns18,Gall18,Phillips25}.

To control systematic uncertainties in the peak magnitudes, we uniformly refit the ASAS-SN $g$-band light curves for all SNe using a suite of subtype-specific templates. The fits are performed in flux space using a $\chi^2$ minimization routine. For normal SNe~Ia, we use the $g$-band flux templates from the Carnegie Supernova Project \citep[CSP;][]{Burns18}, which are extracted using \texttt{SNooPy} \citep{burns11}, and incorporate the dependence of absolute magnitude on $s_{BV}$. The single-band nature of the ASAS-SN light curves precludes color information, leaving the fitted $s_{BV}$ values poorly constrained.

For the peculiar subtypes (91T-like, Iax, 03fg-like, 02es-like, and Ia-CSM), we employ custom templates derived from Gaussian Process fits to the light curves of representative, well-observed events from the literature \citep[e.g.,][]{Krisciunas17,Parrent16,Dimitriadis23,Xi24,Yang23}. Detailed descriptions of these templates and the specific SNe used to construct them are provided in Appendix~\ref{app:lc_temp}. For unclassified SNe, we fit all available templates and select the one that provides the best fit to the peak and the light-curve morphology. To reject outlier data points, we employ a two-pass sigma-clipping procedure: an initial 15$\sigma$ clip followed by a final 5$\sigma$ clip on the residuals of the best-fit model. We determine uncertainties on the best-fit parameters using Monte Carlo methods. 

We compute absolute magnitudes for all SNe using
\begin{equation}
    M_{g} = m_{g} - \mu - A_{g,\text{MW}} - K_g(z),
    \label{eq:absmag}
\end{equation}
where $m_{g}$ is the apparent $g$-band peak magnitude from our light-curve fits, $\mu$ is the distance modulus for the supernova redshift obtained using the \texttt{astropy.cosmology} package \citep{astropy13, astropy18}, and $A_{g,\text{MW}}$ is the Milky Way extinction from \citet{sfd11}. When available, we use redshift-independent distance moduli for $\mu < 32$~mag ($z<0.006$) due to the significance of peculiar velocities at these low redshifts (for more details, see Appendix~\ref{app:z_indep_distmod}). The K-correction, $K_g(z)$, is computed using \texttt{SNooPy} \citep{burns11} with the spectral templates from \citet{hsiao07}. We do not apply a correction for host-galaxy extinction because the single-band ASAS-SN photometry does not provide the color information required to do so robustly. While the light curve fits yield peak apparent magnitudes with statistical uncertainties of $\sim0.05$~mag, the systematic uncertainty in the absolute peak magnitudes is dominated by errors in the distance estimates and host-galaxy extinction.

\begin{figure}
    \centering
    \includegraphics[width=\linewidth]{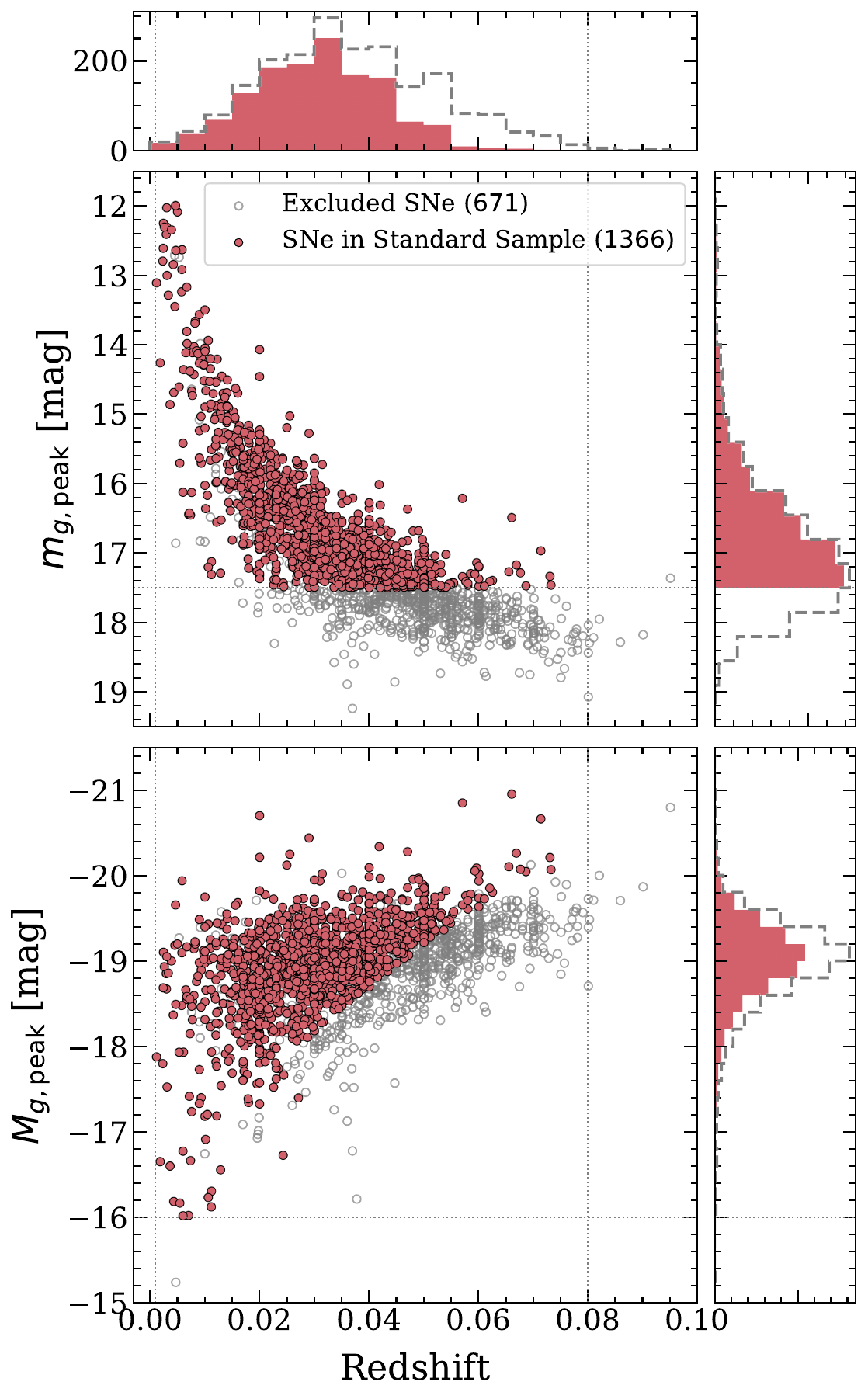}
    \caption{Peak apparent (top) and absolute (bottom) magnitude versus redshift distributions of the 2037 SNe Ia discovered or recovered by ASAS-SN (gray). The filled circles and filled histograms show our $g$-band standard sample of 1366 SNe Ia. Open gray points are the SNe excluded by our standard cuts on peak absolute magnitude $M_{g,\mathrm{peak}}$, Galactic latitude $b$, peak apparent magnitude $m_{g,\mathrm{peak}}$, and redshift $z$, as marked by the gray dotted lines.}
    \label{fig:z_m_hist}
\end{figure}

\begin{table*}
    \centering
    \caption{The $g$-band sample of Type Ia SNe discovered or recovered by ASAS-SN from UTC 2018-01-01 to UTC 2024-12-31.}
    \label{tab:sample}
    \renewcommand{\arraystretch}{1.25}
    \begin{tabular}{l c S[table-format=3.4] S[table-format=-2.4] c c c c c c}
    \toprule
    {IAU Name} & {ASASS-SN Name} & {RA} & {Dec} & {Type} & {$z$} & {$t_{g,\text{peak}}$$^a$} & {$m_{g,\text{peak}}$$^a$} & {$M_{g,\text{peak}}$$^b$} & Disc./Rec.$^c$\\
     & & {[deg]} & {[deg]} & & & {[JD]} & {[mag]} & {[mag]} & \\
    \midrule
    2018eml & ---         & 236.5296 & 29.7335  & Ia           & 0.0316 & 2458343.06 & 16.67 & $-19.08$ & R \\
    2018hme & ASASSN-18yf & 143.9137 & -17.3864 & Ia           & 0.0141 & 2458412.98 & 15.72 & $-18.50$ & D \\
    2019aos & ASASSN-19co & 198.1609 & -30.0234 & Ia-91T-like  & 0.0300 & 2458529.85 & 16.74 & $-19.01$ & D \\
    2019hma & ---         & 157.2444 & 8.4795   & Ia           & 0.0470 & 2458652.47 & 17.55 & $-19.06$ & R \\
    2020eee & ---         & 180.0149 & 10.3723  & Ia           & 0.0710 & 2458927.11 & 18.13 & $-19.39$ & R \\
    2020hvf & ---         & 170.3602 & 3.0147   & Ia-03fg-like & 0.0058 & 2458979.08 & 12.63 & $-19.94$ & R \\
    2021vpv & ---         & 335.5216 & 19.8767  & Ia-91T-like  & 0.0400 & 2459446.02 & 16.90 & $-19.43$ & R \\
    2021wjb & ASASSN-21qf & 300.2080 & -38.5772 & Ia           & 0.0200 & 2459452.06 & 15.79 & $-19.09$ & D \\
    2022eyw & ---         & 190.9999 & 62.3301  & Iax          & 0.0090 & 2459673.76 & 15.63 & $-17.34$ & R \\
    2022yz  & ---         & 35.3532  & 30.1326  & Ia           & 0.0400 & 2459608.76 & 16.84 & $-19.59$ & R \\
    2023vha & ---         & 11.3832  & 31.0076  & Ia           & 0.0820 & 2460246.74 & 17.95 & $-20.00$ & R \\
    2023yma & ---         & 13.0572  & 44.3324  & Ia           & 0.0178 & 2460286.95 & 15.71 & $-19.06$ & R \\
    2024jeg & ---         & 198.4803 & 21.5768  & Ia           & 0.0459 & 2460462.23 & 17.42 & $-19.10$ & R \\
    2024nnu & ASASSN-24ei & 23.7599  & 41.2482  & Ia           & 0.0170 & 2460493.42 & 15.67 & $-18.83$ & D \\
    \bottomrule
    \end{tabular}
    \tablecomments{This table of 2037 SNe~Ia is available in its entirety in a machine-readable form in the online journal. A portion is shown here for guidance regarding its form and content. \\
    $^a$ Peak apparent magnitude and time of peak are derived from refitting ASAS-SN light curves with subtype-specific templates. \\
    $^b$ Peak absolute magnitudes are computed using $m_{g,\text{peak}}$ and Equation~\ref{eq:absmag}. \\
    $^c$ Indicates whether the SN was discovered by ASAS-SN (D) or independently recovered in ASAS-SN data (R).}
\end{table*}

For our standard analysis, we apply several cuts to the full sample of 2037 SNe~Ia to create a statistically robust subset. We use a limiting Galactic latitude of $|b| > 15\degr$, a limiting peak absolute magnitude of $M_{g,\text{peak}} < -16.0$~mag, and a limiting peak apparent magnitude of $m_{g,\text{peak}} < 17.5$~mag, where our survey completeness is well-characterized. These cuts are justified in Section~\ref{subsec:total_rate}. We restrict the redshift range to $0.001 < z < 0.08$. After applying these cuts, our final $g$-band sample for rate calculations consists of 1366 supernovae. Their distribution in $m_{g,\text{peak}}$, $M_{g,\text{peak}}$, and redshift is shown in Figure~\ref{fig:z_m_hist}. The properties of the full $g$-band sample of 2037 SNe~Ia are listed in Table~\ref{tab:sample}.

To leverage the full 11-year baseline of ASAS-SN, we combine this $g$-band sample with the $V$-band sample from Paper~I \citepalias{Desai24}. We apply the same selection criteria to the 2014--2017 data for Galactic latitude, absolute magnitude, and redshift, with the only difference being the apparent magnitude limit ($m_{V,\text{peak}} < 17.0$\,mag versus $m_{g,\text{peak}} < 17.5$\,mag). These cuts contribute an additional 410 SNe~Ia from the $V$-band era. The final combined dataset used for our standard rate analysis therefore consists of 1776 SNe~Ia.

\section{Rate Computations} \label{sec:rates}
We follow the approach of Papers~I \citepalias{Desai24} and II \citep{Pessi25} to compute the volumetric rates, using injection-recovery simulations to estimate the survey completeness as a function of magnitude.

\begin{figure}
    \centering
    \includegraphics[width=\linewidth] {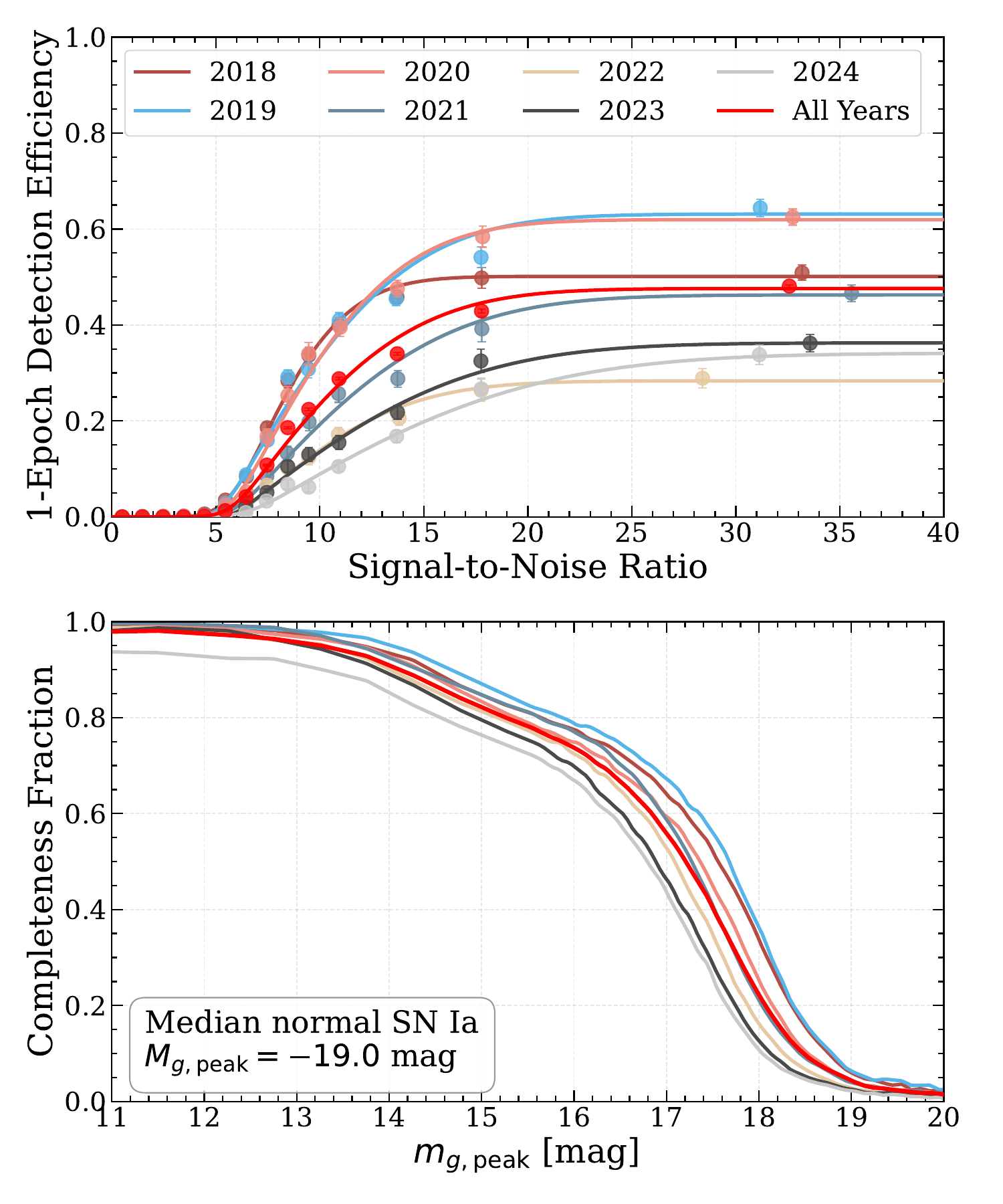}
    \caption{The ASAS-SN $g$-band detection performance.
    \textit{Top:} The probability of flagging a source as a detection in a \textit{single} epoch as a function of SNR for each year. The colored points show binned efficiency estimates from the data, and the solid lines are the best fits from Equation~\ref{eq:efficiency}.
    \textit{Bottom:} The completeness fraction as a function of peak apparent magnitude per year for a typical normal SN~Ia ($M_{g,\text{peak}} = -19.0$\,mag). While the single-epoch efficiency (top) saturates much below unity, the high cadence of ASAS-SN ensures a higher total detection probability for brighter transients (bottom).}
    \label{fig:efficiency}
\end{figure}

Papers~I and II used a simple piecewise function for the probability of detection ($p$) in any given epoch as a function of signal-to-noise ratio (SNR). While this was sufficient for the $V$-band data through 2017, the detection efficiency in later years ($g$-band) varied due to operational changes and resource constraints related to the COVID-19 pandemic and other factors. To account for this, we adopt a more sophisticated parameterization for the detection efficiency from 2018 to 2024. In general, the detection probability is low for low SNR, rises, and then saturates at high SNR. We model this behavior using a cumulative skew-normal distribution with height $H$, center $\mu$, width $\sigma$, and skewness $\alpha$, given by
\begin{equation} \label{eq:efficiency}
    p = \frac{H}{\sigma} \int_{-\infty}^{\text{SNR}} \phi\left(\frac{x-\mu}{\sigma}\right) \left[ 1 + \text{erf}\left( \frac{\alpha(x-\mu)}{\sigma\sqrt{2}} \right) \right] dx
\end{equation}
where $\phi$ is the probability density function of the standard normal distribution.

Based on all $g$-band observations of supernovae (discovered, recovered, and missed), we fit Equation~\ref{eq:efficiency} to the detection efficiency as a function of SNR for each year separately. Figure~\ref{fig:efficiency} (top) shows these fits, and Table~\ref{tab:efficiency} presents the best-fit parameters for each year. Although this represents the probability of detection per observation, the high cadence of ASAS-SN ($1-2$~days in the $g$-band) renders the probability of missing a bright SN negligible. Even if Equation~\ref{eq:efficiency} underestimates $p$ for bright SNe, these objects have many detection trials ($n$), and the overall detection probability, $1-(1-p)^n$, rapidly converges to unity. For example, with $n=5$ trials, the probability of detection for a bright SN is $96.2\%$ for an average $p=0.48$, compared to nearly $100.0\%$ for an idealized $p=0.95$. This total completeness fraction per year is demonstrated for a typical SN~Ia in the bottom panel of Figure~\ref{fig:efficiency}. This completeness includes both weather and seasonal losses.

\begin{table}
\centering
\caption{Yearly and aggregate best-fit parameters for the detection probability defined in Equation~\ref{eq:efficiency}.}
\label{tab:efficiency}
\renewcommand{\arraystretch}{1.25} 
\setlength{\tabcolsep}{8pt}      
\begin{tabular}{c|cccc}
\toprule
Year & \textbf{$H$} & \textbf{$\mu$} & \textbf{$\sigma$} & \textbf{$\alpha$} \\
\midrule
2018    & 0.50 & 5.68 & 3.95  & 3.60  \\
2019    & 0.63 & 5.23 & 6.68  & 10.13 \\
2020    & 0.62 & 5.71 & 5.87  & 6.76  \\
2021    & 0.46 & 5.85 & 7.58  & 8.82  \\
2022    & 0.28 & 6.06 & 5.98  & 7.37  \\
2023    & 0.36 & 5.82 & 8.61  & 18.41 \\
2024    & 0.34 & 6.49 & 10.53 & 9.72  \\
\midrule
$2018 - 2024$ & 0.48 & 5.69 & 6.50  & 7.71  \\
\bottomrule
\end{tabular}
\end{table}

\begin{figure}
    \centering
    \includegraphics[width=\linewidth]{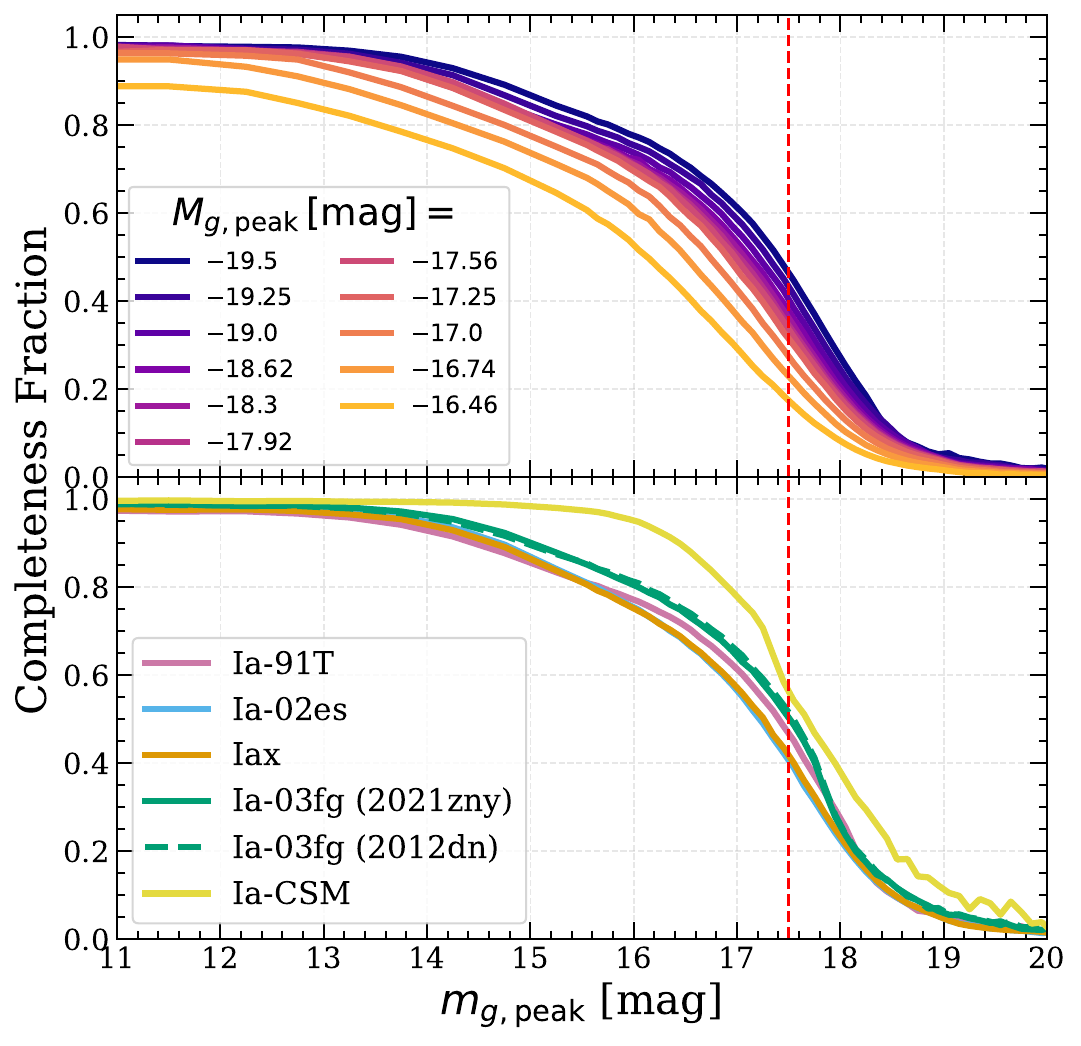}
    \caption{\textit{Top}: Completeness fraction ($F_1$) for all years combined for the templates used for normal SNe~Ia. \textit{Bottom}: Completeness fraction ($F_1$) for all years combined for other subtypes of SNe~Ia. The dashed red line marks the standard limiting magnitude of $g=17.5\,\mathrm{mag}$.}
    \label{fig:completeness}
\end{figure}

We performed injection-recovery simulations by drawing $N_{\mathrm{LC}} = 100{,}000$ random ASAS-SN $g$-band light curves. For each trial, we injected a synthetic SN light curve based on the templates for each subtype and peak absolute magnitude. For normal SNe~Ia, we used the CSP-II templates \citep{Burns18}, which relate the peak absolute magnitude to the color-stretch parameter $s_{BV}$. Each synthetic observation was then flagged as a ``detection" or ``non-detection" based on the probability from Equation~\ref{eq:efficiency}. A trial SN was considered detected if at least one epoch was flagged as a detection. We conducted a total of $M=100\,N_{\mathrm{LC}}$ trials, resulting in a detection fraction of $F_1=N/M$ divided by year, where $N$ is the number of detected trials. The resulting completeness fractions for all years combined are shown for each of our SN~Ia templates in Figure~\ref{fig:completeness}.

As in Papers~I and II, we introduce a second correction factor, $F_2(M_{\mathrm{\textit{g},peak}}) = V(z_{\mathrm{lim}}(M_{\mathrm{\textit{g},peak}})) / V(z_{\mathrm{max}})$, to account for the varying maximum redshift, $z_{\mathrm{lim}}$, used for different absolute magnitudes in the simulations. This factor normalizes the detection volume to that of our maximum redshift $z_{\mathrm{max}}$.

\begin{figure*}
    \centering
    \includegraphics[width=\textwidth]{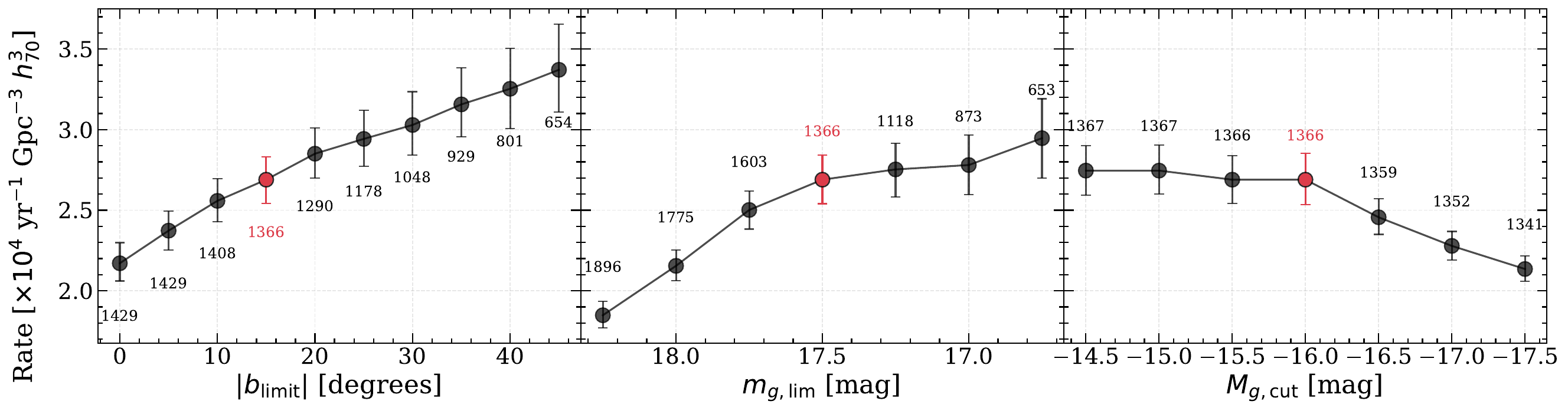}
    \caption{\textit{Left}: Rate as a function of the Galactic latitude cut. \textit{Middle}: Rate as a function of the limiting apparent magnitude. \textit{Right}: Rate as a function of faintest peak absolute magnitude cut. In all three panels, the numbers are the number of SNe in each bin. The red points in each panel are the values used for our standard sample.}
    \label{fig:rate_vs_cuts}
\end{figure*}

Finally, since the spectroscopic completeness varies by up to $\sim20\%$ with year and apparent magnitude, we adopt a third correction factor, $F_3$. This factor is the fraction of SNe with spectroscopic classifications in a given year and apparent magnitude bin. It assumes that the intrinsic distribution of unclassified SNe is the same as that of classified SNe within a given magnitude bin, which is a reasonable approximation for this work. We discuss spectroscopic completeness further in Appendix~\ref{app:spec_comp}. The final statistical weight for the $i^{\mathrm{th}}$ observed SN is $w_{i,g} = \left(F_{1,i}\, F_{2,i}\, F_{3,i}\right)^{-1}$. The volumetric SN rate is then
\begin{equation}
    R_g = \frac{\sum_{i=1}^{N} w_{i,g}}{V\, \Delta t_g \, \left(1-\sin{b_{\mathrm{lim}}}\right)} ,
    \label{eq:rate}
\end{equation}
where $\Delta t_g = 7.0\, \mathrm{yr}$ is the time span from UTC 2018-01-01 to UTC 2024-12-31, $V = \frac{4}{3}\pi \left( d_{\mathrm{max}}^3 - d_{\mathrm{min}}^3 \right)$ is the total comoving volume, and $(1-\sin{b_\mathrm{lim}})$ corrects for our Galactic latitude limit.

To maximize the statistical precision of our measurement, we combine the $V$-band (2014--2017) and $g$-band (2018--2024) samples, extending the effective survey duration to $\Delta t_{\mathrm{tot}} = 11.0$\,yr. In doing so, we use the CSP DR3 light curves \citep{Krisciunas17} to approximate a relation between $m_{g,\text{peak}} - m_{V,\text{peak}}$ and $M_{V,\text{peak}}$ to convert the $V$-band magnitudes to $g$-band. This method is described in Appendix~\ref{app:g_band_convert} and has a scatter of 0.12~mag. While intrinsic color differences and host-galaxy extinction effects introduce systematic uncertainty at the magnitude cutoff, this effect is negligible for the rates of subtypes or between luminosity bins compared to the Poisson errors. To ensure the samples span consistent volumes, other than the different limiting apparent magnitudes ($m_{g,\text{peak}} < 17.5$~mag, $m_{V,\text{peak}} < 17.0$~mag), we apply identical constraints to both datasets: $z_{\min}=0.001$, $z_{\max}=0.08$, and a peak absolute magnitude limit of $M_{g, \mathrm{peak}} < -16.0$\,mag. The final combined rate is calculated by summing the weights from both survey phases over the total probed volume-time:
\begin{equation}
    R_{V+g} = \frac{\sum_{j=1}^{N_V} w_{j,V} + \sum_{i=1}^{N_g} w_{i,g}}{V \Delta t_{\mathrm{tot}} (1 - \sin b_{\lim})} ,
    \label{eq:rate_combined}
\end{equation}
where $w_{i,g}$ are the weights derived in this work for the $N_g=1366$ $g$-band SNe, and $w_{j,V}$ are the weights derived in \citetalias{Desai24} for the $N_V=410$ $V$-band SNe. By combining these datasets, we achieve the sample size necessary to robustly constrain the rates of rare subtypes where Poisson errors are the limiting factor.

\section{Volumetric Rates} \label{sec:vol_rates}
\subsection{Total Rate} \label{subsec:total_rate}
Figure~\ref{fig:rate_vs_cuts} illustrates the sensitivity of the total volumetric rate to our primary selection criteria. We adopt a Galactic latitude cut of $|b| > 15\degr$ to maximize the sample size while avoiding the high extinction and stellar crowding of the Galactic plane (left panel). The middle panel shows the stability of the rate as a function of the limiting apparent magnitude. The rate remains consistent within uncertainties for $m_{g,\text{lim}} \leq 17.5$~mag but declines at fainter magnitudes, suggesting that our completeness corrections are imperfectly accounting for the rapid drop in detection efficiency. Consequently, we adopt $m_{g,\text{lim}} = 17.5$~mag as the limit for the standard sample.

The choice of the absolute magnitude cut ($M_{g,\text{cut}}$) significantly impacts both the calculated rate and its precision (right panel). Extending the limit to include intrinsically faint SNe (e.g., pushing $M_{g,\text{cut}}$ from $-17.0$ to $-15.0$~mag) increases the total volumetric rate, indicating the presence of a significant low-luminosity population. However, because these events are difficult to detect, they require large completeness corrections ($w_{i,g}$). The combination of their significant contribution to the total number and their high statistical weights results in a disproportionate inflation of the uncertainty. This trade-off was also observed in the $V$-band analysis of Paper~I \citepalias{Desai24}. To balance the inclusion of fainter populations with statistical robustness, we adopt $M_{g,\text{cut}} = -16.0$~mag for our standard sample.

For our standard $g$-band sample of 1366 SNe~Ia ($M_{g,\text{peak}} < -16.0$~mag, $m_{g,\text{peak}} < 17.5$~mag, $|b| > 15\degr$, and $0.001 < z < 0.08$), using Equation~\ref{eq:rate}, we measure a total volumetric rate of
\begin{equation}
R_{\text{tot},g} = (2.69 \pm 0.14) \times 10^{4} \, \mathrm{yr}^{-1}\, \mathrm{Gpc}^{-3}\, h^{3}_{70},
\end{equation}
at a median redshift of $z_{\text{med}} = 0.03$. The fractional uncertainty of $5.2\%$ exceeds the Poisson expectation ($2.7\%$) due to the non-uniform weighting required to correct for completeness driven by the subset of SNe with low detection probabilities and high weights. However, thanks to the $\sim 3.4$ times larger sample size of the $g$-band survey, there is a significant improvement in precision over the $V$-band measurement \citepalias[9\% uncertainty;][]{Desai24}.

By combining the $V$- and $g$-band samples (Equation~\ref{eq:rate_combined}), we leverage the full statistical power of the 11-year ASAS-SN baseline. Using the 1776 SNe~Ia with $M_{g,\text{peak}} < -16.0$~mag, $m_{g,\text{peak}} < 17.5$~mag, $m_{V,\text{peak}} < 17.0$~mag$, |b| > 15\degr$, and $0.001 < z < 0.08$, we find a final ASAS-SN SN~Ia rate of
\begin{equation} \label{eq:tot_rate}
R_{\mathrm{tot}} = (2.55 \pm 0.12) \times 10^{4} \, \mathrm{yr}^{-1}\, \mathrm{Gpc}^{-3}\, h^{3}_{70}
\end{equation}
at a median redshift of $z=0.029$.
With a total uncertainty of only $4.7\%$, this is the most precise measurement of the local volumetric SN~Ia rate to date.

While we will discuss rates as a function of galaxy properties in later papers, we can rephrase Equation~\ref{eq:tot_rate} in units roughly representing the rate in $L_*$ galaxies like the Milky Way. The density scale of $L_*$ galaxies is $n_* \approx 0.01\,h^3\,\mathrm{Mpc^{-3}}$ \citep[e.g.,][]{Blanton03, Bell03} and $h=0.7$, so the rate per $L_*$ galaxy per century is approximately
\begin{equation} \label{eq:L_star_rate}
    R_{\mathrm{tot}}n_*^{-1} = 0.74 \,(L_*\,\mathrm{galaxy)^{-1}}\,\mathrm{century^{-1}}.
\end{equation}
This is consistent with estimates of the SN~Ia rate in the Galaxy of  $1.4^{+1.4}_{-0.8}\,\mathrm{century^{-1}}$ \citep[e.g.,][]{Adams13}. Equivalently, a rate of one SN~Ia per $L_*$ galaxy per century is $3.4 \times 10^{4} \, \mathrm{yr}^{-1}\, \mathrm{Gpc}^{-3}\, h^{3}_{70}$.

\begin{figure*}
    \centering
    \includegraphics[width=\textwidth]{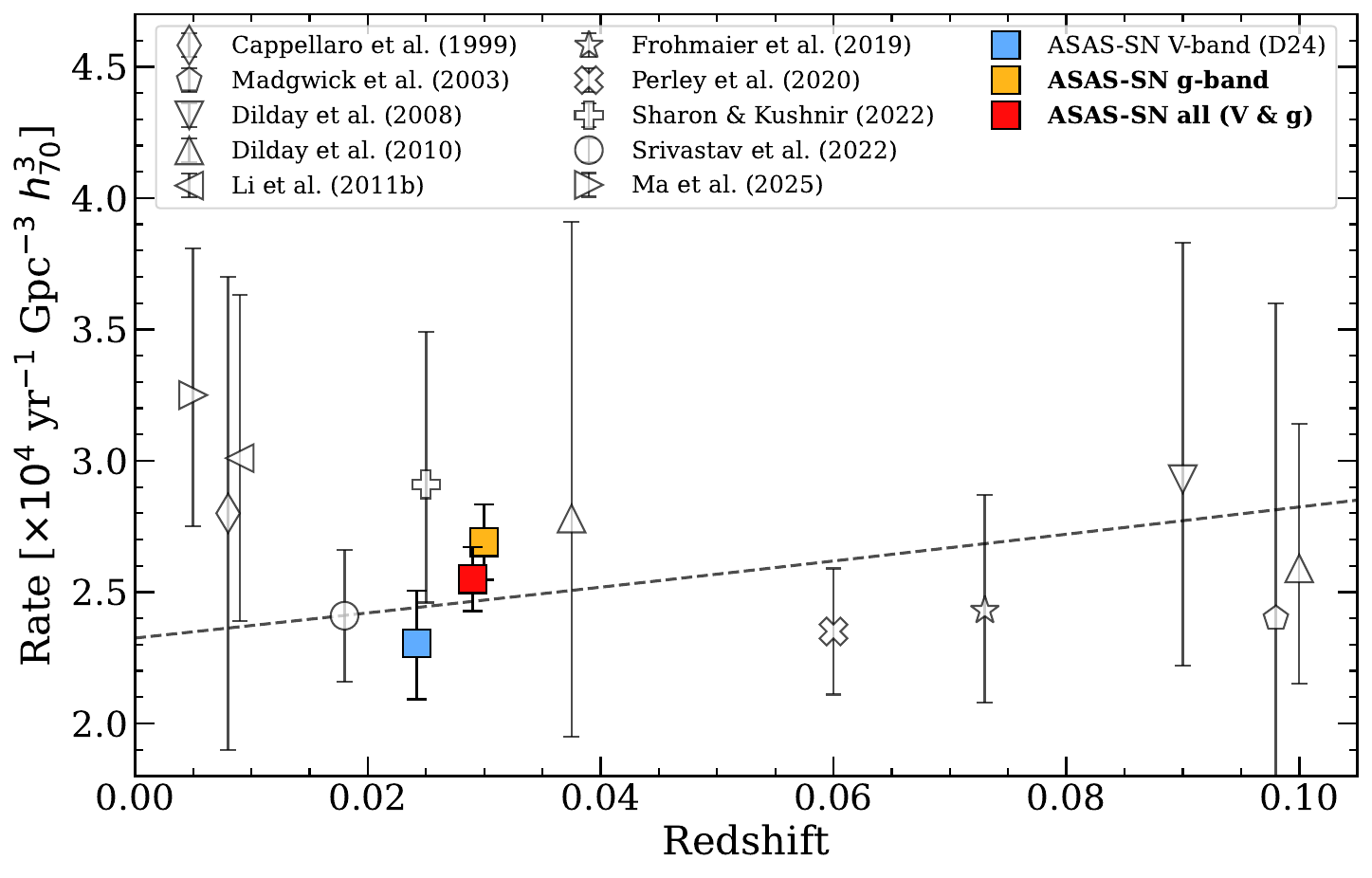}
    \caption{Volumetric SN Ia rate as a function of redshift $z$. ASAS-SN rates from $V$-band \citepalias{Desai24} and $g$-band, separately, and the final rate from combining the entire ASAS-SN sample are shown as filled squares at the median redshift of each sample ($M_{g,\mathrm{peak}} < -16.0$~mag). Rates from other studies are shown as open symbols \citep{cappellaro99,Madgwick03,Dilday08,dilday10,li11b,frohmaier19,perley20,sharon-kushnir22,Srivastav22,Ma25b}. The dashed line shows the power-law derived by \citet{dilday10} for all SNe~Ia (including all subtypes), scaling as $(1+z)^{\alpha}$ with $\alpha = 2.04^{+0.90}_{-0.89}$.}
    \label{fig:rates_vs_z}
\end{figure*}

Figure~\ref{fig:rates_vs_z} places the total rate in the context of previous studies. When interpreting these comparisons, the differing absolute magnitude limits of each survey must be considered. While \citet{frohmaier19} restricted their sample to `cosmologically useful' SNe~Ia ($M < -18.0$ mag), \citet{perley20} reached $-16.5$ mag, and some volume-limited studies integrated as faint as $-15.0$ or $-14.0$ mag \citep{Srivastav22,Ma25b}. Because the SN~Ia luminosity function remains relatively flat down to $M \approx -16$ mag (see Section~\ref{sec:LF}), the total integrated rate is sensitive to the choice of this cutoff. Consequently, the value we report is formally the rate for SNe~Ia with $M_{g,\mathrm{peak}} < -16.0$ mag, rather than the rate for all SNe that could be spectroscopically classified as SNe~Ia regardless of their absolute magnitude.

With this definition, our rate from Equation~\ref{eq:tot_rate} is statistically consistent with measurements from other untargeted surveys. These include PTF \citep[$(2.43 \pm 0.29) \times 10^{4} \, \mathrm{yr}^{-1}\, \mathrm{Gpc}^{-3}\, h^{3}_{70}$;][]{frohmaier19}, the ZTF Bright Transient Survey \citep[$(2.35 \pm 0.24) \times 10^{4} \, \mathrm{yr}^{-1}\, \mathrm{Gpc}^{-3}\, h^{3}_{70}$;][]{perley20}, and ATLAS \citep[$(2.41 \pm 0.25) \times 10^{4} \, \mathrm{yr}^{-1}\, \mathrm{Gpc}^{-3}\, h^{3}_{70}$;][]{Srivastav22}. The volume-limited ZTF rate from \citet{sharon-kushnir22} is slightly higher, although statistically consistent; their analysis includes a correction for host-galaxy extinction, which will increase the estimated rate. While broadly consistent with these studies, the ASAS-SN measurement benefits from reduced systematic uncertainties. Unlike previous efforts, we explicitly model the survey completeness as a function of both peak absolute magnitude and light-curve shape, and over observing seasons, using subtype-specific templates to accurately characterize the selection function across the full diversity of the SN~Ia population, down to the faintest luminosities.

In contrast, our rate is lower than estimates from older galaxy-targeted surveys \citep[e.g.,][]{cappellaro99, li11b}, though the large uncertainties in those works maintain statistical consistency. We observe a mild tension with the recent measurement of $3.25_{-0.50}^{+0.56} \times 10^{4} \, \mathrm{yr}^{-1}\, \mathrm{Gpc}^{-3}\, h^{3}_{70}$ by \citet{Ma25b}; this likely stems from cosmic variance within their smaller volume ($D < 40$\,Mpc) and methodological differences between characterizing the selection function of a heterogeneous compilation versus a single, uniform survey. Finally, our local rate measurement agrees with the expected $(1+z)^\alpha$ evolution derived from higher-redshift samples \citep[e.g.,][]{dilday10}.

\subsection{Subtype Rates} \label{subsec:subtype_rates}
A primary motivation for this work is to leverage the statistical power of the combined 11-year ASAS-SN sample to robustly measure the volumetric rates of SN~Ia subtypes. These rates and fractions are presented in Table~\ref{tab:vol_rates_subtypes}. 

\begin{table}
\centering
\caption{Volumetric rates and relative fractions for SN~Ia subtypes using the combined $V$+$g$ dataset. The primary rows show the standard sample ($M_{g,\mathrm{peak}} < -16.0$~mag). Rows in italics show the values for the fainter absolute magnitude cut ($M_{g,\mathrm{peak}} < -15.0$~mag) where they differ; values not listed remain unchanged. All fractions are relative to the total rate for the specific sample.}
\label{tab:vol_rates_subtypes}
\renewcommand{\arraystretch}{1.5} 
\setlength{\tabcolsep}{4.4pt} 
\begin{tabular}{l | c c c}
\toprule
Subtype & $N$ & Rate $[\mathrm{yr}^{-1}\, \mathrm{Gpc}^{-3}\, h^{3}_{70}]$ & Fraction [\%] \\
\midrule
Ia-norm+other & 1642  & $(2.36 \pm 0.11) \times 10^4$  & $92.7 \pm 1.9$ \\
\quad ($\mathit{M < -15}$) & -- & -- & $\mathit{91.4 \pm 1.7}$ \\
Iax           & 9     & $800 \pm 400$                  & $3.0 \pm 1.4$ \\
\quad ($\mathit{M < -15}$) & $\mathit{10}$ & $\mathit{(1.1 \pm 0.5) \times 10^3}$ & $\mathit{4.3 \pm 1.8}$ \\
Ia-91T        & 108   & $850 \pm 100$                  & $3.3 \pm 0.4$ \\
Ia-02es       & 4     & $210 \pm 130$                  & $0.8 \pm 0.5$ \\
Ia-03fg       & 8     & $32 \pm 12$                    & $0.12 \pm 0.05$ \\
Ia-CSM        & 5     & $9 \pm 4$                      & $0.036 \pm 0.017$ \\
\midrule
\textbf{Total} & \textbf{1776} & $\mathbf{(2.55 \pm 0.12) \times 10^4}$ & \textbf{100}\\
\textit{\textbf{\quad ($\boldsymbol{M} < \boldsymbol{-15}$)}} & \textit{\textbf{1777}} & $(\textit{\textbf{2.58}} \pm \textit{\textbf{0.12}}) \times \textit{\textbf{10}}^{\textit{\textbf{4}}}$ & \textit{\textbf{100}} \\
\bottomrule
\end{tabular}
\end{table}

The ``Ia-norm+other'' category, which includes normal SNe~Ia, transitional events, subluminous 91bg-like events, and SNe lacking a specific subtype classification, dominates the population, accounting for $(92.7 \pm 1.9)\%$ of the total rate in our standard sample. While we broadly categorize 91bg-like events as part of the normal distribution, we estimate a lower limit of $R_{\text{91bg}} \ge (1.7 \pm 0.4) \times 10^3 \, \mathrm{yr}^{-1} \, \mathrm{Gpc}^{-3} \, h_{70}^3$ ($\ge 6.7\%$ of total SN Ia rate) for the rate of this specific spectroscopic subclass using only those events explicitly classified as 91bg-like on TNS. This estimate is notably lower than the $\sim 15-16\%$ fractions reported by previous studies \citep[][]{li11a, Ma25a}. However it is inherently a lower limit, as it is restricted to SNe with spectra of sufficient quality showing a strong Ti~II absorption feature; additional candidates likely remain hidden within the Ia-norm+other population. It is also likely that the galaxy-targeted strategy of the LOSS sample \citep{li11a} inherently biases the 91bg-like rate because it focused on more massive hosts, and fast-declining, 91bg-like SNe are preferentially found in massive early-type galaxies \citep[e.g.,][]{Hamuy00, Neill09, Barkhudaryan19, Rigault20}.

With 108 91T-like SNe observed in our standard sample, we measure a volumetric rate of $R_{\text{91T}} = (850 \pm 100) \, \mathrm{yr}^{-1} \, \mathrm{Gpc}^{-3} \, h_{70}^3$ ($3.3 \pm 0.4\%$). With a more precise measurement, our observed fraction for this subtype is still consistent with the $V$-band analysis in Paper~I \citepalias[$4\%$;][]{Desai24} and the $5.4^{+3.6}_{-3.8}\%$ fraction found in the recent compilation by \citet{Ma25a}.  It is also statistically consistent ($1.5\sigma$) with the higher fraction of $9.4_{-4.7}^{+5.9}\%$ reported for the volume-limited LOSS sample \citep{li11a}, owing to their large uncertainties.

We find that SNe~Iax (02cx-like) comprise $(3.0 \pm 1.4)\%$ of the total rate in our standard sample ($M_{g,\mathrm{peak}} < -16.0$ mag). Extending the limit to $M_{g,\mathrm{peak}} < -15.0$ mag adds one SN~Iax to the sample (and no other subtypes), raising the Iax fraction to $(4.3 \pm 1.8)\%$ and marginally increasing the total SN~Ia volumetric rate to $R_{\text{tot}} = (2.58 \pm 0.12) \times 10^4 \, \mathrm{yr}^{-1} \, \mathrm{Gpc}^{-3} \, h_{70}^3$. These fractions are consistent with the $5.4^{+4.7}_{-3.3}\%$ and $6.8^{+4.0}_{-3.4}\%$ reported by \citet{li11a} and \citet{Ma25a}, respectively. Our results are also statistically consistent with the higher fraction of $15^{+17}_{-9}\%$ estimated by \citet{Srivastav22} using a volume-limited sample from ATLAS. While the uncertainties on their measurements are large, \citet{Srivastav22} find that the Iax population is dominated by low-luminosity events ($M \gtrsim -16$ mag), which alone account for $\sim 12\%$ of the total SN~Ia rate in their analysis. Because these subluminous events largely fall below our standard absolute magnitude cut ($M < -16$ mag) and sensitivity limits, our measured rate likely represents a lower bound. If the Iax LF does not turnover rapidly at $M_{g,\mathrm{peak}} > -15.0$~mag, these faint explosions could contribute significantly more to the total SN~Ia volumetric rate.

The rarest subtype are the SNe showing interaction with circumstellar material (CSM). We find a rate for SNe~Ia-CSM of $R_{\text{Ia-CSM}} = (9 \pm 4) \, \mathrm{yr}^{-1}\, \mathrm{Gpc}^{-3}\, h^{3}_{70}$, corresponding to only $(0.036 \pm 0.017)\%$ of the total SN~Ia rate. This rate is consistent with the $V$-band measurement of $R_{\text{Ia-CSM}} = 10 \pm 7 \, \mathrm{yr}^{-1}\, \mathrm{Gpc}^{-3}\, h_{70}^{3}$ \citepalias{Desai24} and with the low fractions ($0.02-0.2\%$) found in the ZTF Bright Transient Survey \citep{Sharma23}, but much better constrained. The presence of dense, hydrogen-rich CSM has been attributed to various progenitor scenarios, such as mergers within a common envelope or winds from non-degenerate companions; however, the specific mechanism required to produce the often massive CSM inferred for these events remains debated \citep[e.g.,][]{Hamuy03, Livio03, Lundqvist13, Meng17}. Regardless of the exact progenitor channel(s), our measurement confirms the extreme rarity of these SNe~Ia-CSM, a conclusion reinforced by the strict limits on late-time CSM interaction derived from ultraviolet observations \citep[e.g.,][]{Graham19, Dubay22}. This provides strong demographic evidence that this specific evolutionary pathway contributes less than $0.05\%$ to the total population of thermonuclear explosions in the local universe.

We provide new constraints on the rates of two of the most peculiar and physically enigmatic SN~Ia subtypes: the subluminous 02es-like and the luminous 03fg-like SNe. We find $R_{\text{02es}} = (210 \pm 130)\, \mathrm{yr}^{-1}\, \mathrm{Gpc}^{-3}\, h^{3}_{70}$ and $R_{\text{03fg}} = (32 \pm 12) \, \mathrm{yr}^{-1}\, \mathrm{Gpc}^{-3}\, h^{3}_{70}$. While the 03fg-like rate is consistent with the $V$-band measurement from Paper~I, the rate for 02es-like events is marginally higher. This increase is likely driven by the $g$-band survey's deeper limiting magnitude ($g \lesssim 17.5$~mag vs $V \lesssim 17.0$~mag), allowing a lower-luminosity limit, and improved spectroscopic classification efficiency for these rare events.

These two subtypes have recently been linked observationally. \citet{Hoogendam24} showed that 02es-like and 03fg-like SNe are the \textit{only} subtypes to exhibit non-monotonic bumps in their rising light curves \citep[e.g.,][]{Cao15, Miller20, Burke21, Jiang21, Srivastav23_2022ilv, Dimitriadis23, AbreuPaniagua25}. Based on these shared properties, recent studies have theorized that these events might represent different ends of a physical continuum, perhaps defined by progenitor core mass \citep[e.g.,][]{Ashall21,Hoogendam24}. Regardless of the specific physical driver, our rates provide constraints on such scenarios.

We find that 02es-like SNe are significantly more common than 03fg-like SNe, with a rate ratio of $R_{\text{02es}} / R_{\text{03fg}} = 7 \pm 5$. Since 02es-like events are faint and extend to the faintest absolute magnitudes, this ratio is really a lower limit. If these subtypes indeed form a continuum, the underlying progenitor channel must strongly favor the production of lower-luminosity explosions. This provides a demographic constraint for progenitor models, such as violent white dwarf mergers \citep[e.g.,][]{Pakmor10, Kromer16} or core-degenerate scenarios \citep[e.g.,][]{Kashi11, Ashall21}, that attempt to explain their shared properties of a C-rich envelope and unusual rising light curves~\citep{Hoogendam24}.

\section{Luminosity Functions} \label{sec:LF}
\begin{figure}
    \centering
    \includegraphics[width=\linewidth]{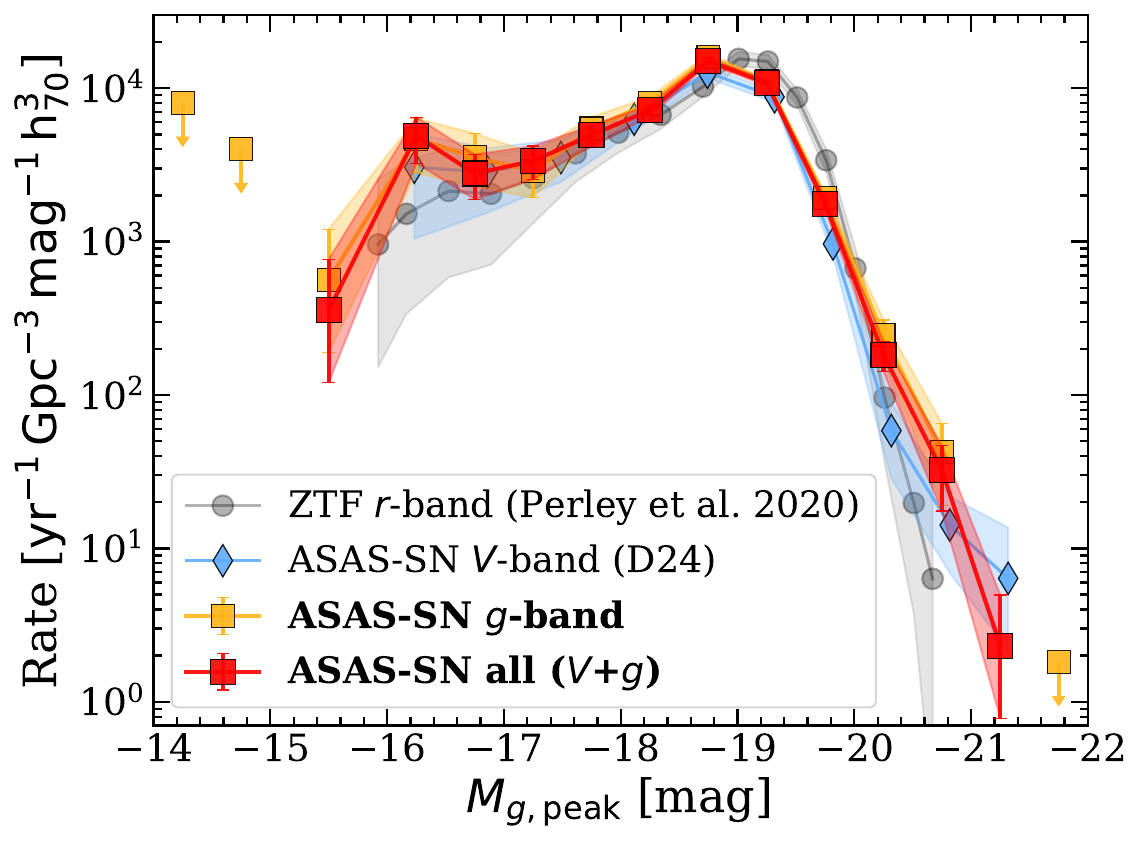}
    \caption{Comparison of the total SN~Ia luminosity functions derived from the combined ASAS-SN $V$+$g$ sample (red squares) and $g$-band only sample (golden squares) with the V-band LF from Paper~I \citepalias[blue diamonds;][]{Desai24} and the ZTF r-band LF from \citet{perley20} (grey circles). The combined sample provides tighter constraints, particularly at the faint end.}
    \label{fig:rates_vs_M_compare}
\end{figure}

We construct the luminosity functions (LFs) for SNe~Ia and their subtypes by leveraging the full statistical power of the combined $V$+$g$ ASAS-SN dataset. This 11-year baseline enables the most detailed demographic study of SN~Ia subtypes in the local universe to date. The resulting LFs, corrected only for Galactic extinction, are presented in Figure~\ref{fig:rates_vs_M_compare}.

Following the methodology of \citetalias{Desai24}, we divide the SNe in our combined sample (Section~\ref{sec:sample}) into $0.5$~mag bins and calculate the differential volumetric rate using Equation~\ref{eq:rate_combined}. 
For the two least luminous and the one most luminous magnitude bins containing no detected events, we calculate upper limits using a Monte Carlo approach. We simulate $10^5$ artificial sources uniformly distributed in comoving volume and magnitude within the bin limits. By convolving these simulated sources with our survey completeness function, we determine the mean detection probability and the resulting effective survey volume. The reported limits correspond to the volumetric rate that would yield a Poisson expectation of 1.15 detected events ($1\sigma$ confidence) given this effective volume-time product.

The total SN~Ia LF (red squares in Figure~\ref{fig:rates_vs_M_compare}) exhibits the characteristic shape observed in previous studies, peaking at $M_{g,\mathrm{peak}} \approx -19$~mag. Figure~\ref{fig:rates_vs_M_compare} compares our results to the $V$-band-only LF from Paper~I \citepalias{Desai24} and the ZTF $r$-band LF from \citet{perley20}. For a direct comparison, we transformed these distributions into the $g$ band following the method detailed in Appendix~\ref{app:g_band_convert}; this procedure applies the necessary offsets to the absolute magnitudes and scales the rates to account for differences in bin width. The results are broadly consistent within the uncertainties across the full luminosity range, though our $g$-band and combined LFs extend to lower luminosities. While the data suggest a downturn faintward of $M_{g,\mathrm{peak}} \approx -16$~mag, we note that the faintest bin contains only a single SN and is heavily reliant on volume corrections. Nevertheless, the LF must eventually turn over.

We briefly consider the implications for host-galaxy extinction similar to Paper I \citepalias{Desai24}. The agreement between the $g$- and $V$-band LFs, with no discernible magnitude offset, is likely due to the significant overlap in their filter response functions. By comparing these distributions to the ZTF $r$-band LF from \citet{perley20}, we estimate a mean host extinction of $E(V-r)_{\mathrm{host}} \approx E(g-r)_{\mathrm{host}} \approx 0.2$~mag. It is important to note, however, that this value does not account for very highly extincted SNe that are largely excluded from magnitude-limited optical surveys.

\begin{figure*}
    \centering
    \includegraphics[width=\textwidth]{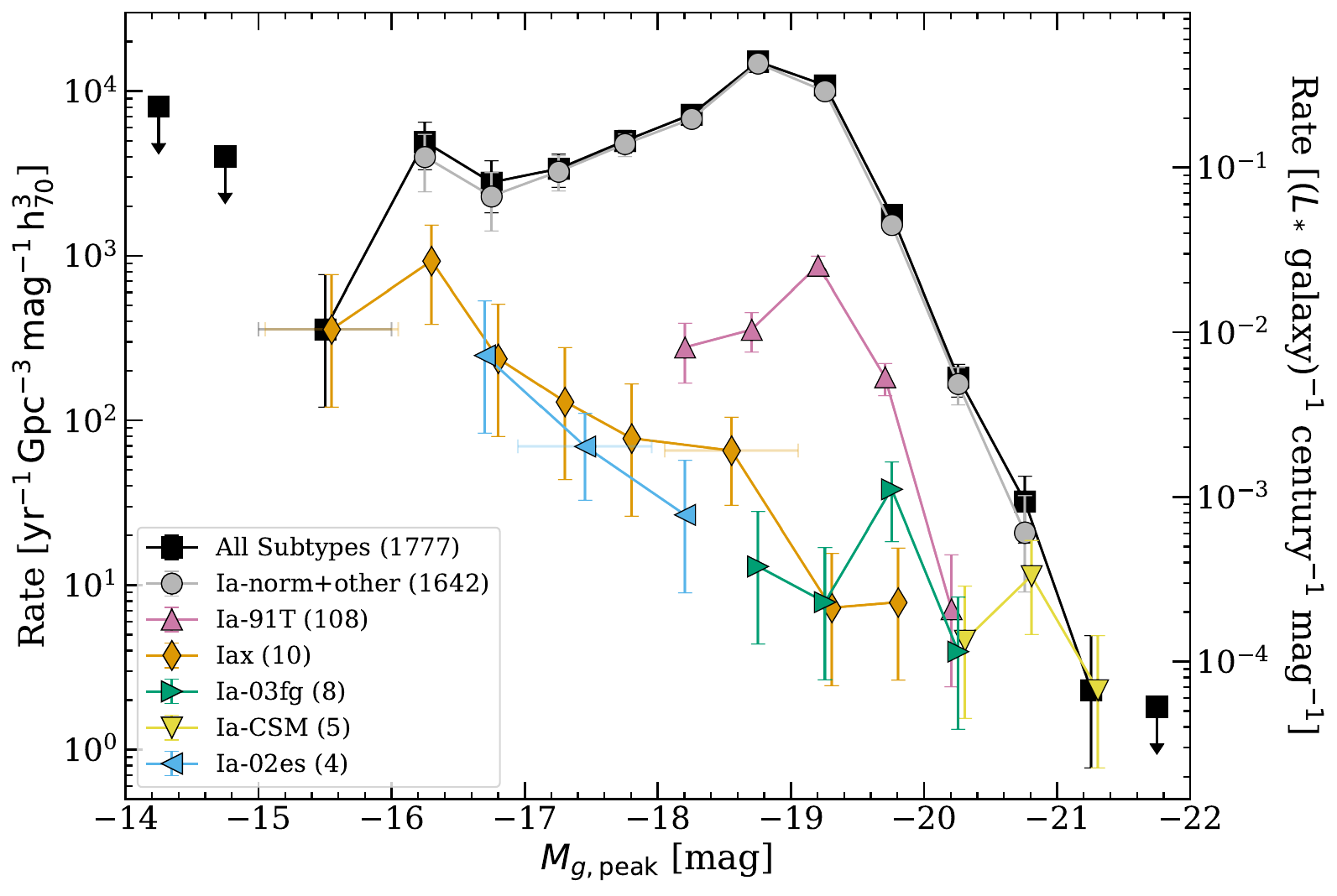}
    \caption{The $g$-band luminosity functions (LF) for SN~Ia subtypes derived from the combined ASAS-SN $V$- and $g$-band samples. The total LF is shown in black, while LFs for individual subtypes are color-coded as indicated in the legend (sample sizes in parentheses). The points with downward arrows are $1\sigma$ upper limits. For clarity, the bin centers for SN~Ia-91T and SN~Ia-02es are shifted by $+0.05$~mag, and those for SN~Iax and SN~Ia-CSM by $-0.05$~mag. The standard bin width is 0.5~mag and the bins marked with horizontal bars have a width of 1.0~mag. The right axis shows the rate normalized to an $L_*$ galaxy per century per mag assuming $n_*=0.01\,h^3\,\mathrm{Mpc^{-3}}$ \citep[e.g.,][]{Blanton03, Bell03}. The tabulated data for these LFs, including the upper limits for all subtypes, are available in the ancillary materials.}
    \label{fig:LF_subtypes}
\end{figure*}

Figure~\ref{fig:LF_subtypes} presents the subtype-specific LFs with the data and upper limits given in an ancillary file. To provide physical context, the right axis shows the rate normalized to that for $L_*$ (Milky Way-like) galaxy per century, as done for Equation~\ref{eq:L_star_rate}.

The luminosity function of 91T-like SNe exhibits a spread of $\sim 2$ mag. This broad distribution is consistent with the preference of 91T-like SNe for star-forming host galaxies \citep[e.g.,][]{Howell01,Sullivan10,Phillips24}, where the progenitors are often subject to significant host-galaxy extinction of up to $E(B-V)_{\mathrm{host}} \approx 0.4$ mag \citep{Phillips22}. Additionally, the likely inclusion of 99aa-like and shallow-silicon events contributes to the width of the 91T-like distribution, as these subtypes are intrinsically fainter than pure 91T-like events \citep[e.g.,][]{Phillips22}.

The SNe~Iax LF is notably broad, spanning nearly five magnitudes from $M_{g,\text{peak}} \approx -20$~mag and rising towards our faint limit of $-15$ mag. This diversity aligns with recent evidence suggesting that SNe~Iax comprise two distinct populations based on absolute magnitude and decline rates: a luminous group ($M_{\mathrm{peak}} \lesssim -16$ mag) and a low-luminosity group ($M_{\mathrm{peak}} \gtrsim -16$ mag) \citep[e.g.,][]{Magee16, Srivastav22, Singh23}. 
The low-luminosity population includes some extremely faint events such as SNe~2008ha, 2010ae, 2019gsc, 2020kyg, and 2021fcg, all with $M_{\mathrm{peak}} > -15$ mag \citep{Foley09, Stritzinger14, Srivastav20, Srivastav22, Karambelkar21}. While our measured Iax fraction ($4.3 \pm 1.8\%$) is statistically consistent with volume-limited estimates \citep{li11a,Srivastav22,Ma25a}, the rising LF suggests we may be missing a significant low-luminosity tail of the distribution. \citet{Srivastav22} estimate that a significant fraction of SNe~Iax are fainter than $M_{\mathrm{peak}} = -16$ mag; given that many of these events fall below our magnitude cut, the true intrinsic rate of SNe~Iax is likely higher. This luminosity distribution is consistent with pure deflagration models of Chandrasekhar-mass WDs, which predict a diverse range of outcomes from weak partial deflagrations to near-complete disruptions \citep{Fink14, Jha17}.

Finally, we provide the first robust LFs for the peculiar 02es-like and 03fg-like subtypes from a systematic survey. The 03fg-like LF spans $M_{\mathrm{peak}} \approx -20.5$ to $-18.5$ mag, whereas the 02es-like LF occupies a distinctly fainter regime from $M_{\mathrm{peak}} \approx -18.5$ down to $-16.5$ mag. When viewed together, these subtypes populate a continuous range of luminosities divided roughly at $M_{\mathrm{peak}} \approx -18.5$ mag. This observed luminosity continuum is consistent with theoretical frameworks where these subtypes arise from a single progenitor channel governed by a physical parameter, such as core mass \citep[e.g.,][]{Ashall21,Hoogendam24}. The difference in peak magnitude likely correlates with effective temperature, naturally explaining the spectroscopic differences between the cooler, Ti~II-rich 02es-like events and the hotter 03fg-like events. Furthermore, the combined 02es-like and 03fg-like LF is heavily skewed towards the faint end, reinforcing the implication that any common progenitor channel must strongly favor lower-luminosity explosions. Additionally, if the 02es-like LF continues to rise faintward of $M_{g,peak} = -16.5$~mag, similar to the behavior discussed for SNe~Iax, then our reported rate represents only a lower bound on a potentially much larger population of low-luminosity transients.

\section{Summary \& Conclusion} \label{sec:summary}
In this third study of SN rates in ASAS-SN, we have analyzed the combined 11-year sample of SNe~Ia discovered and recovered by ASAS-SN between 2014 and 2024. By integrating the high spectroscopic completeness of the $V$-band era (Paper I; \citetalias{Desai24}) with the depth and statistics of the $g$-band era, we provide the most statistically rigorous benchmark for the local SN~Ia population to date. Our main conclusions are as follows:

\begin{itemize}
    \item We measure a total volumetric SN~Ia rate for our standard sample ($M_{g,\mathrm{peak}} < -16.0$~mag) of $R_{\mathrm{tot}} = (2.55 \pm 0.12) \times 10^4 \, \mathrm{yr}^{-1} \, \mathrm{Gpc}^{-3} \, h_{70}^3$. This result is consistent with previous untargeted surveys \citep[e.g.,][]{frohmaier19,perley20,Srivastav22} but offers a significant improvement in precision ($4.7\%$ total uncertainty) due to the extended baseline and uniform selection function of ASAS-SN. This measurement serves as a robust anchor for determining the redshift evolution of the SN~Ia rate.

    \item We provide robust volumetric rates for the SN~Ia subtypes. The ``Ia-norm+other'' category dominates, accounting for $(92.7 \pm 1.9)\%$ of the total rate. For the luminous 91T-like SNe, we measure a rate of $(850 \pm 100) \text{ yr}^{-1} \text{ Gpc}^{-3} h_{70}^3$ ($3.3 \pm 0.4\%$ of the total). SNe Iax have a rate of $(4.3 \pm 1.8)\%$ of the total rate when considering SNe down to $M_{g,\mathrm{peak}} < -15.0$~mag. However, because the SNe~Iax LF rises toward lower luminosity, this value likely represents a lower bound on their true population.

    \item We find that the subluminous 02es-like SNe ($210 \pm 130 \text{ yr}^{-1} \text{ Gpc}^{-3} h_{70}^3$) are significantly more common than the luminous 03fg-like SNe ($32 \pm 12 \text{ yr}^{-1} \text{ Gpc}^{-3} h_{70}^3$). The ratio of their rates is $R_{\mathrm{02es}}/R_{\mathrm{03fg}} = 7 \pm 5$. This large ratio and their LFs place strong demographic constraints on theoretical models that propose a physical continuum for these two subtypes \citep[e.g.,][]{Ashall21,Hoogendam24}, implying that any common progenitor channel must strongly favor the production of lower-luminosity explosions. If there are any undetected or unclassified 02es-like SNe below $M_{g,\mathrm{peak}} = -16.5$~mag, then the true volumetric rate of these events will be higher than measured here.

    \item We confirm the extreme rarity of SNe~Ia-CSM associated with dense circumstellar material. We measure a rate for SNe Ia-CSM of $(9 \pm 4) \text{ yr}^{-1} \text{ Gpc}^{-3} h_{70}^3$, corresponding to only $(0.036 \pm 0.017)\%$ of the total SNe~Ia rate. This implies that the evolutionary pathways producing massive CSM shells around SN Ia progenitors are exceptionally rare in the local universe.

    \item The combined luminosity functions reveal immense diversity in the SNe~Ia population. SNe~Ia-CSM appear only in the most luminous bins. The 02es-like and 03fg-like subtypes occupy distinct luminosity ranges that together form a potential continuum. The SNe~Iax distribution is extremely broad, extending from $M_{g,\mathrm{peak}} \approx -20$~mag down to the limit of $-15.0$~mag. 
\end{itemize}

The total SN~Ia LF is broad and spans more than five magnitudes. This has significant implications for the physical parameters of the explosions. If the peak luminosity simply scales linearly with the mass of synthesized radioactive nickel \citep[$L_{\mathrm{peak}} \propto M_{^{56}\mathrm{Ni}}$;][]{Arnett82}, this implies that the nickel yields of SNe~Ia vary by a factor of nearly 100. This range is significantly larger than the factor of $\sim 10$ typically invoked in theoretical models \citep[e.g.,][]{Piro14}. Furthermore, to the extent that nickel mass correlates with the total kinetic energy of the explosion, the resulting SN remnants must originate from events spanning two orders of magnitude in energy.

Finally, we discuss the low-luminosity end of the LF. While the total LF shows a tentative downturn fainter than $M_{g,\mathrm{peak}} \approx -16$~mag, we find no statistical evidence for a decline in the rates of subluminous subtypes, such as SNe~Iax and 02es-like SNe. The true contribution of these low-luminosity transients depends heavily on the behavior of the LF beyond our current limits. We consider three scenarios. First, if the rate declines rapidly, our current measurements of the SN~Ia rate and subtype fractions are robust. Second, if the LF remains flat at $\approx 10^3\,\mathrm{yr}^{-1}\,\mathrm{Gpc}^{-3}\,\mathrm{mag}^{-1}$ past $-16$~mag, the population would need to extend down to $9$~mag to account for 50\% of the total SN~Ia rate, and down to $-3$~mag to account for 25\%. This would be nonphysical for SNe. Third, if the rate continues to rise with the slope observed for SNe~Iax ($\log_{10}(R_{\mathrm{Iax}}) \propto 0.475 M_{g,\mathrm{peak}}$) these events could be more abundant than their higher-luminosity counterparts, reaching 50\% of the total rate if such a trend extends to $\sim-12.5$~mag. A SN at $-12.5$~mag is only detectable within $10$~Mpc for ASAS-SN's $17.5$~mag limit. We likely underestimate the abundance of these low-luminosity events because candidates are frequently recognized as intrinsically faint only post-peak, bypassing triggers designed for studying rising, normal SNe~Ia. To fully characterize the physics of thermonuclear explosions, we encourage time-domain surveys with deeper limiting magnitudes to prioritize the nearby volume, specifically targeting transients fainter than $-16$~mag to constrain this faint population.

The next paper in this series will extend this 11-year analysis to the core-collapse supernova population. Subsequent work will then leverage these combined subtype-specific samples to determine rates as a function of host-galaxy properties. By isolating the environmental dependencies of distinct populations, we aim to map specific evolutionary pathways to their observed outcomes. This demographic analysis is essential for establishing a physical link between the progenitor systems, their host environments, and the resulting explosion mechanisms.

\section*{Acknowledgments}
We thank Federica Chiti for helpful discussions. 

The Shappee group at the University of Hawai`i is supported with funds from NSF (grant AST-2407205) and NASA (grants HST-GO-17087, 80NSSC24K0521, 80NSSC24K0490, 80NSSC23K1431). CSK and KZS are supported by NSF grants AST-2307385 and AST-2407206. ASAS-SN is funded by Gordon and Betty Moore Foundation grants GBMF5490 and GBMF10501 and the Alfred P. Sloan Foundation grant G-202114192. JFB is supported by NSF grant PHY-2310018. SD is supported by the National Natural Science Foundation of China (Grant No. 12133005). JL is supported by NSF-2206523.

\bibliography{references}
\bibliographystyle{mnras}

\appendix
\counterwithin{figure}{section}
\counterwithin{table}{section}
\renewcommand{\thefigure}{\thesection\arabic{figure}}
\renewcommand{\thetable}{\thesection\arabic{table}}
\renewcommand{\theHequation}{\thesection\arabic{equation}}
\renewcommand{\theHfigure}{\thesection\arabic{figure}}
\renewcommand{\theHtable}{\thesection\arabic{table}}

\section{Light Curve Templates} \label{app:lc_temp}
To accurately model our sample, we require templates that cover the full diversity of SN~Ia morphologies. While standard stretch-based templates are sufficient for normal SNe~Ia, they often fail to capture the unique evolutionary features of peculiar subtypes.

\begin{figure}
    \centering
    \includegraphics[width=\linewidth]{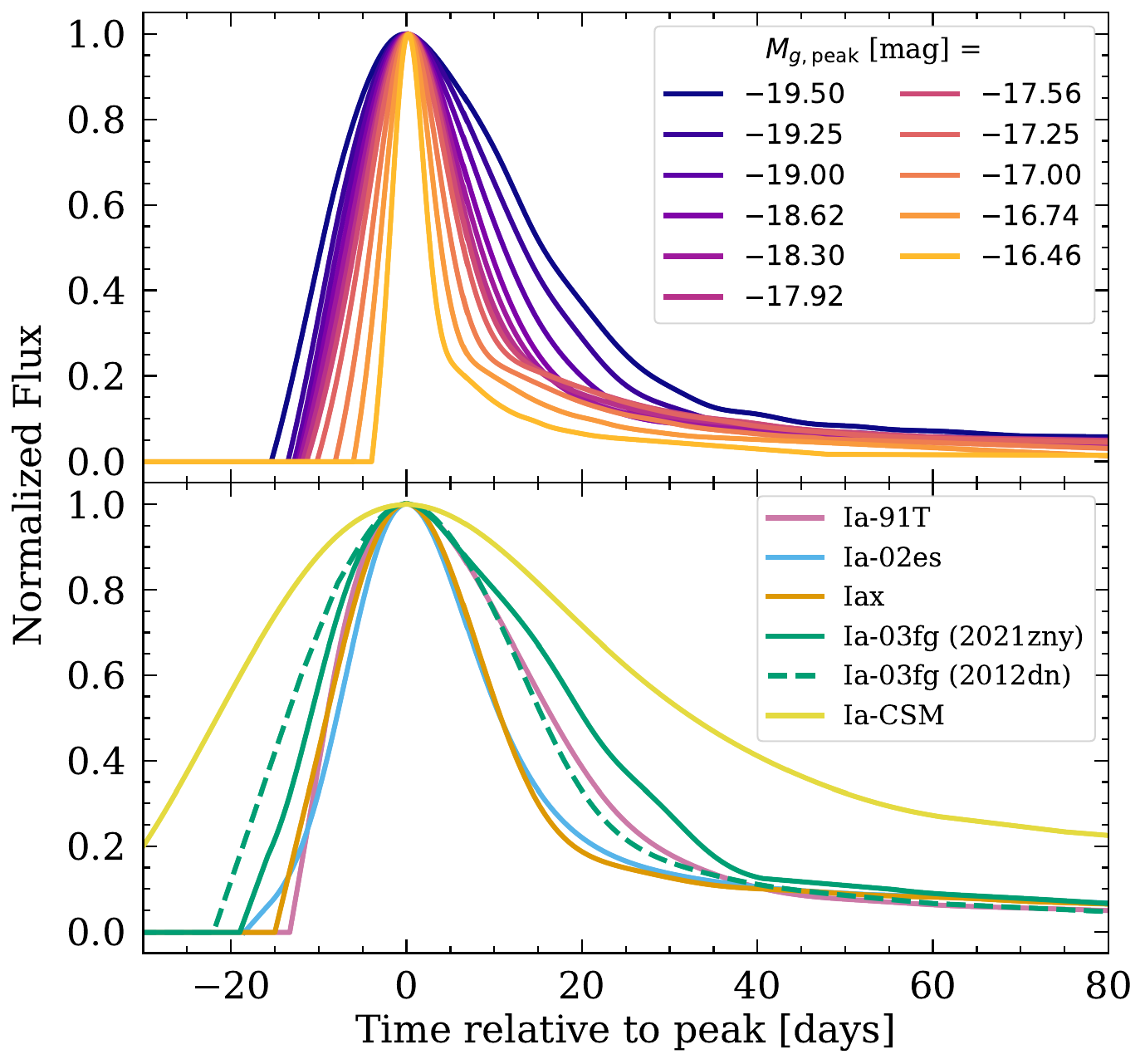}
    \caption{Normalized $g$-band flux templates, aligned at the time of maximum light ($t=0$). \textit{Top:} The family of templates used for normal SNe~Ia. The legend indicates the peak absolute magnitude corresponding to the template shape, illustrating the stretch-luminosity relationship where brighter SNe (dark blue) are broader than fainter ones (yellow). \textit{Bottom:} Custom templates for the subtypes. For 03fg-like SNe, we use multiple templates to capture the range of observed rise times and decline rates.}
    \label{fig:templates}
\end{figure}

For normal (including 91bg-like and transitional) SNe~Ia, we utilize the $g$-band flux templates from the Carnegie Supernova Project (CSP-II; \citealt{Burns18}), accessed via \textsc{SNooPy} \citep{burns11}. These templates, shown in the top panel of Figure~\ref{fig:templates}, effectively parameterize the light curve shape using the color-stretch parameter, $s_{BV}$, which correlates with peak luminosity.

For the peculiar subtypes, we constructed custom templates using Gaussian Process regression on well-sampled light curves of representative objects (Figure~\ref{fig:templates}, bottom panel):
\begin{itemize}
    \item 91T-like: We combined CSP DR3 $g$-band photometry of SNe~2004gu, 2005M, 2005eq, and 2007S \citep{Krisciunas17} to create a master template that captures the broad, slowly declining morphology typical of this class.
    \item Iax: We adopted the CSP light curve of the prototype SN~2005hk \citep{Krisciunas17} as the fiducial template. We acknowledge, however, that this subtype likely exhibits a diversity of light curve shapes not fully captured by a single template.
    \item 02es-like: We generated a template from the high-cadence ASAS-SN light curve of SN~2022vqz \citep{Xi24}, which captures the fast evolution characteristic of this subtype.
    \item 03fg-like: Given the photometric diversity of this subtype, we employ two distinct templates covering the range of observed shapes: one based on SN~2012dn \citep{Parrent16} and another on SN~2021zny \citep{Dimitriadis23}.
    \item Ia-CSM: We constructed a template using the ASAS-SN $g$-band light curves of SN~2018evt and SN~2022erq. These are the only two confirmed Ia-CSM events in our $g$-band sample, and the template exhibits the extreme width expected from interaction-powered emission.
\end{itemize}

\section{Distance Moduli} \label{app:z_indep_distmod}
\begin{figure}
    \centering
    \includegraphics[width=\linewidth]{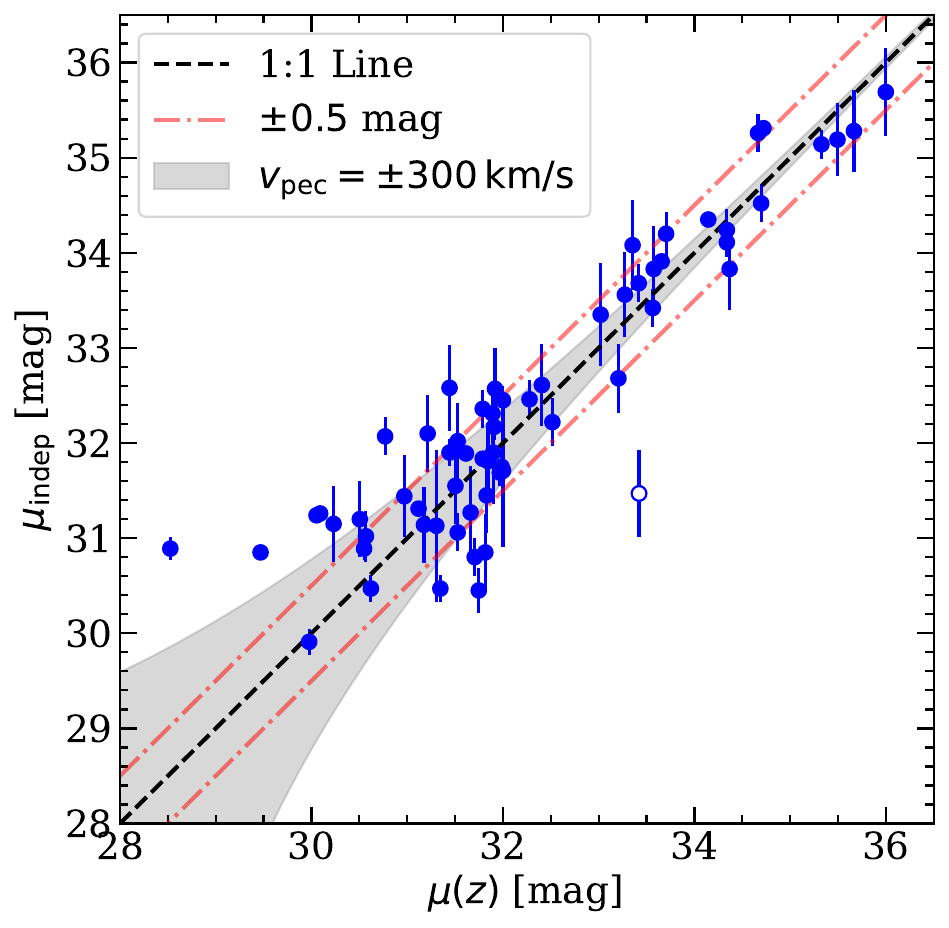}
    \caption{Comparison of redshift-based distance moduli ($\mu(z)$) and redshift-independent distance moduli ($\mu_{\mathrm{indep}}$). The solid black dashed line indicates the 1:1 relation. The gray shaded region represents the uncertainty introduced by a peculiar velocity of $v_{\mathrm{pec}} = \pm 300$ km s$^{-1}$. The red dot-dashed lines mark a $\pm 0.5$ mag deviation, corresponding to the bin size of our luminosity functions. SN~2018fpm (open circle) is an outlier with a likely incorrect literature distance.}
    \label{fig:distmod}
\end{figure}

\begin{table}
    \centering
    \caption{Redshift-independent distance moduli for $z < 0.006$ sample}
    \label{tab:z_indep_distances}
    \renewcommand{\arraystretch}{1.25}
    \resizebox{\columnwidth}{!}{%
    \begin{tabular}{l c c c l}
    \toprule
    SN Name          & Type        & $z$    & $\mu_{\mathrm{indep}}$ & Source \\
                     &             &        & [mag]                  &        \\
    \midrule
    2024gy           & Ia          & 0.0012 & 30.89                  & \citet{Li26} \\
    2021qvv          & 91bg-like   & 0.0018 & 30.85                  & \citet{Graur23} \\
    2020ukx          & Ia          & 0.0023 & 29.91                  & \citet{Tully13} \\
    2021J            & Ia          & 0.0024 & 31.24                  & \citet{Gallego-Cano22} \\
    2020nlb          & Ia          & 0.0024 & 31.26                  & \citet{Sand21} \\
    2020rcq          & Ia          & 0.0026 & 31.15                  & \citet{Tully88} \\
    2020nvb          & Ia          & 0.0029 & 31.20                  & \citet{Tully88} \\
    2020ue           & Ia          & 0.0030 & 30.89                  & \citet{Tully13} \\
    2022pul          & 03fg-like   & 0.0030 & 31.02                  & \citet{Siebert24} \\
    2018pv           & Ia          & 0.0031 & 30.47                  & \citet{Tully13} \\
    2021hiz          & Ia          & 0.0033 & 32.07                  & \citet{Tully13} \\
    2020qxp          & 02es-like   & 0.0036 & 31.44                  & \citet{Hoeflich21} \\
    2021pit          & Ia          & 0.0039 & 31.31                  & \citet{Riess16} \\
    2017cbv          & Ia          & 0.0040 & 31.14                  & \citet{Hosseinzadeh17} \\
    2015bp           & 91bg-like   & 0.0041 & 32.10                  & \citet{Wyatt21} \\
    2021smj          & Ia          & 0.0042 & 31.13                  & \citet{Tully88} \\
    2018imd          & Ia          & 0.0043 & 30.47                  & \citet{Tully13} \\
    ASASSN-14gh      & Ia          & 0.0044 & \nodata                & \nodata \\
    2024inv          & Ia          & 0.0045 & \nodata                & \nodata \\
    2019np           & Ia          & 0.0045 & 32.58                  & \citet{Tully16} \\
    2016coj          & Ia          & 0.0045 & 31.90                  & \citet{Zheng17} \\
    2024muv          & Ia          & 0.0046 & 31.55                  & \citet{Karachentsev13} \\
    2022hrs          & Ia          & 0.0047 & 31.06                  & \citet{Tully13} \\
    2023zgx          & Iax         & 0.0047 & 32.02                  & \citet{Tully88} \\
    2015F            & Ia          & 0.0049 & 31.89                  & \citet{Im15} \\
    2021aefx         & Ia          & 0.0050 & 31.27                  & \citet{Ashall22} \\
    ASASSN-14lp      & Ia          & 0.0051 & 30.80                  & \citet{Shappee16} \\
    2014dt           & Iax         & 0.0052 & 30.45                  & \citet{Foley16} \\
    2024xal          & Ia          & 0.0053 & 32.36                  & \citet{Tully13} \\
    2018gv           & Ia          & 0.0053 & 31.84                  & \citet{Yang20} \\
    2017bzc          & Ia          & 0.0054 & 30.85                  & \citet{Russell02} \\
    2017fzw          & 91bg-like   & 0.0054 & 31.45                  & \citet{Graham22} \\
    2023vyl          & Ia          & 0.0054 & 31.81                  & \citet{Tully88} \\
    2017drh          & Ia          & 0.0056 & 32.31                  & \citet{Hoogendam25pxl} \\
    M OT J1149$^{*}$ & Ia          & 0.0056 & 31.90                  & \citet{Tully16} \\
    2014bv           & Ia          & 0.0056 & 32.17                  & \citet{Tully13} \\
    2017igf          & Ia          & 0.0056 & 32.57                  & \citet{Sorce14} \\
    2022ffv          & Ia          & 0.0058 & 31.69                  & \citet{Tully13} \\
    2018aoz          & Ia          & 0.0058 & 31.75                  & \citet{Ni22} \\
    2020hvf          & 03fg-like   & 0.0058 & 32.45                  & \citet{Jiang21} \\
    ASASSN-15us      & 02es-like   & 0.0058 & 31.71                  & \citet{Tully88} \\
    \bottomrule
    \end{tabular}%
    }
    \tablecomments{$^{*}$Full name: MASTER OT J114925.48-050713.8.}
\end{table}

While redshifts provide reliable distances in the Hubble flow, the peculiar velocities of nearby galaxies introduce significant scatter. To mitigate this, we adopt redshift-independent distances for our lowest redshift events. Figure~\ref{fig:distmod} compares the distance modulus derived from the redshift, $\mu(z)$, against the redshift-independent distance modulus, $\mu_{\mathrm{indep}}$, for the nearby SNe. The gray band illustrates the impact of a $\pm 300$ km s$^{-1}$ peculiar velocity on the Hubble flow distance. As the redshift decreases, this uncertainty grows, exceeding $\pm 0.5$ mag (red dashed lines) at approximately $\mu(z)  \approx 32$ mag ($z \approx 0.006$). Therefore, for SNe with $z < 0.006$, we use redshift-independent distances from the NASA/IPAC Extragalactic Database (NED)\footnote{\url{https://ned.ipac.caltech.edu/}} or recent literature, when available.

We list the distances and their sources in Table~\ref{tab:z_indep_distances}. Our search yielded redshift-independent distance moduli for 39 of the 41 SNe in this regime. One object, SN~2018fpm, deviates significantly from the 1:1 relation at $\mu(z) \approx 33.4$ mag. This discrepancy likely arises from an erroneous Tully-Fisher distance measurement in the literature \citep{Theureau07}.

\section{Spectroscopic Completeness} \label{app:spec_comp}
\begin{figure*}
    \centering
    \includegraphics[width=\textwidth]{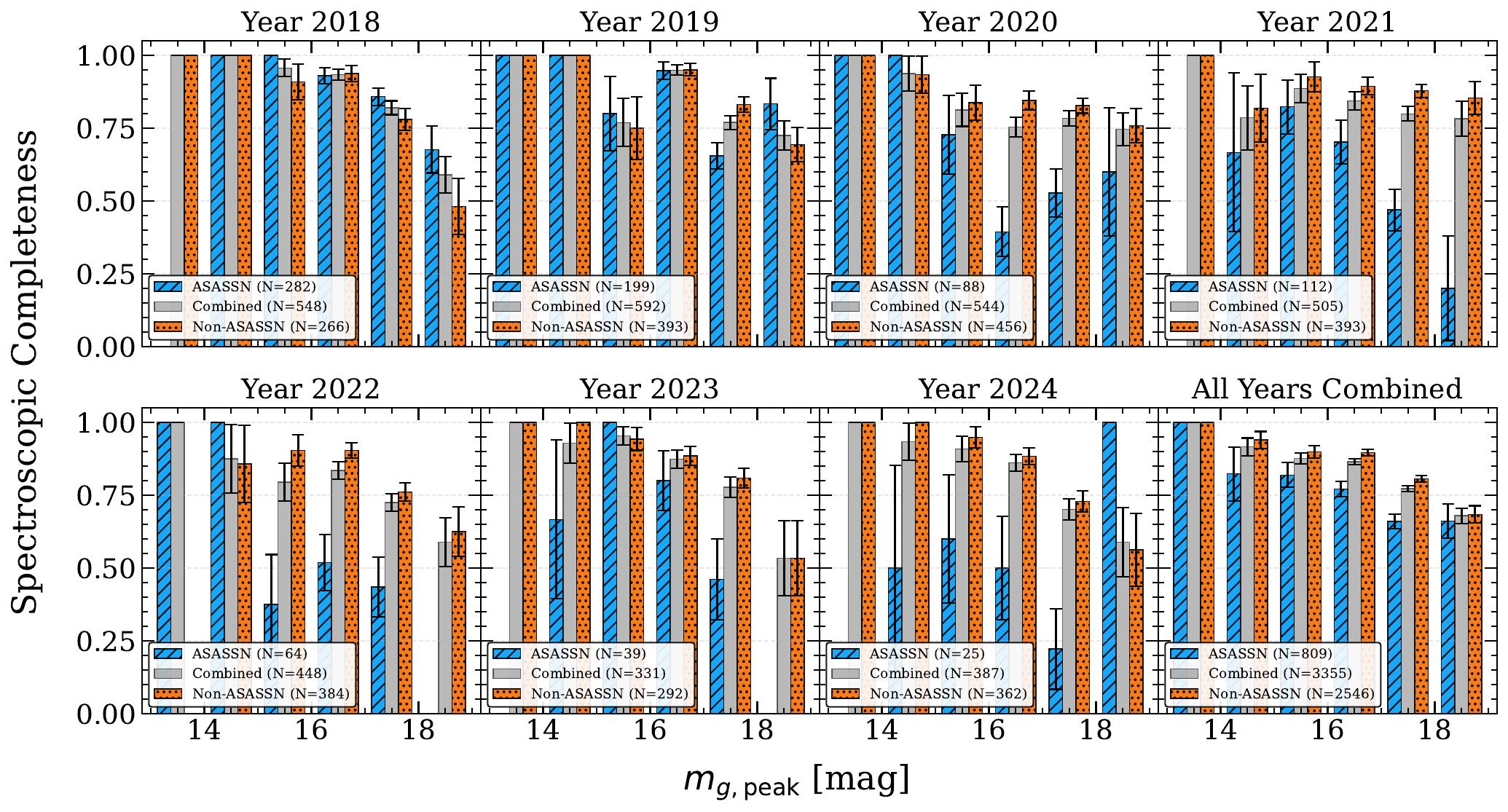}
    \caption{Spectroscopic completeness as a function of peak apparent $g$-band magnitude ($m_{g,\mathrm{peak}}$) and year. The final panel shows the distribution for the full sample. In each plot, the completeness is shown separately for transients discovered by ASAS-SN (blue), those discovered by other surveys but recovered by ASAS-SN (orange), and the combined sample (gray). Error bars represent the binomial uncertainty in each magnitude bin.}
    \label{fig:spec_comp_per_year}
\end{figure*} 
While the $V$-band survey (2014--2017) achieved a nearly spectroscopically complete sample \citep{Holoien17a, Holoien17b, Holoien17c, Holoien19}, the g-band survey (2018--2024) covers a longer baseline and reaches fainter magnitudes, resulting in a larger fraction of unclassified candidates. To incorporate these objects into our rate calculations, we include a spectroscopic completeness correction factor, $F_{3}$. We draw our sample of unclassified candidates from the ASAS-SN discovered or recovered transients, selecting objects that satisfy three criteria: (1) a light curve morphology consistent with a supernova, (2) proximity to a potential host galaxy, and (3) the lack of a stellar counterpart in Gaia, Pan-STARRS, or TESS catalogs.

We bin both the classified SNe~Ia and the unclassified candidates by year and peak apparent magnitude ($m_{g,\mathrm{peak}}$) in 1-mag bins. We define the spectroscopic completeness as
\begin{equation}
    F_3(y, m) = \frac{N_{\mathrm{classified}}(y, m)}{N_{\mathrm{total}}(y, m)},
\end{equation}
where $N_{\mathrm{classified}}$ is the number of spectroscopically confirmed SNe and $N_{\mathrm{total}}$ is the sum of classified SNe and unclassified candidates in a given year ($y$) and magnitude bin ($m$).

Figure~\ref{fig:spec_comp_per_year} presents the spectroscopic completeness as a function of peak magnitude for each year of the $g$-band survey. The completeness drops significantly for fainter objects ($m_{g,\mathrm{peak}} > 17$~mag). This decline is time-dependent and stems from multiple factors. Issues following 
the COVID-19 pandemic disrupted classification resources starting in 2020, and as of 2022 June 1, ASAS-SN ceased soliciting confirmation images from amateur astronomers. This operational change reduced the survey's ``aggressiveness'' in confirming faint or ambiguous candidates, lowering the completeness for fainter candidates in the 2022--2024 datasets compared to earlier years.

\begin{figure}
    \centering
    \includegraphics[width=\linewidth]{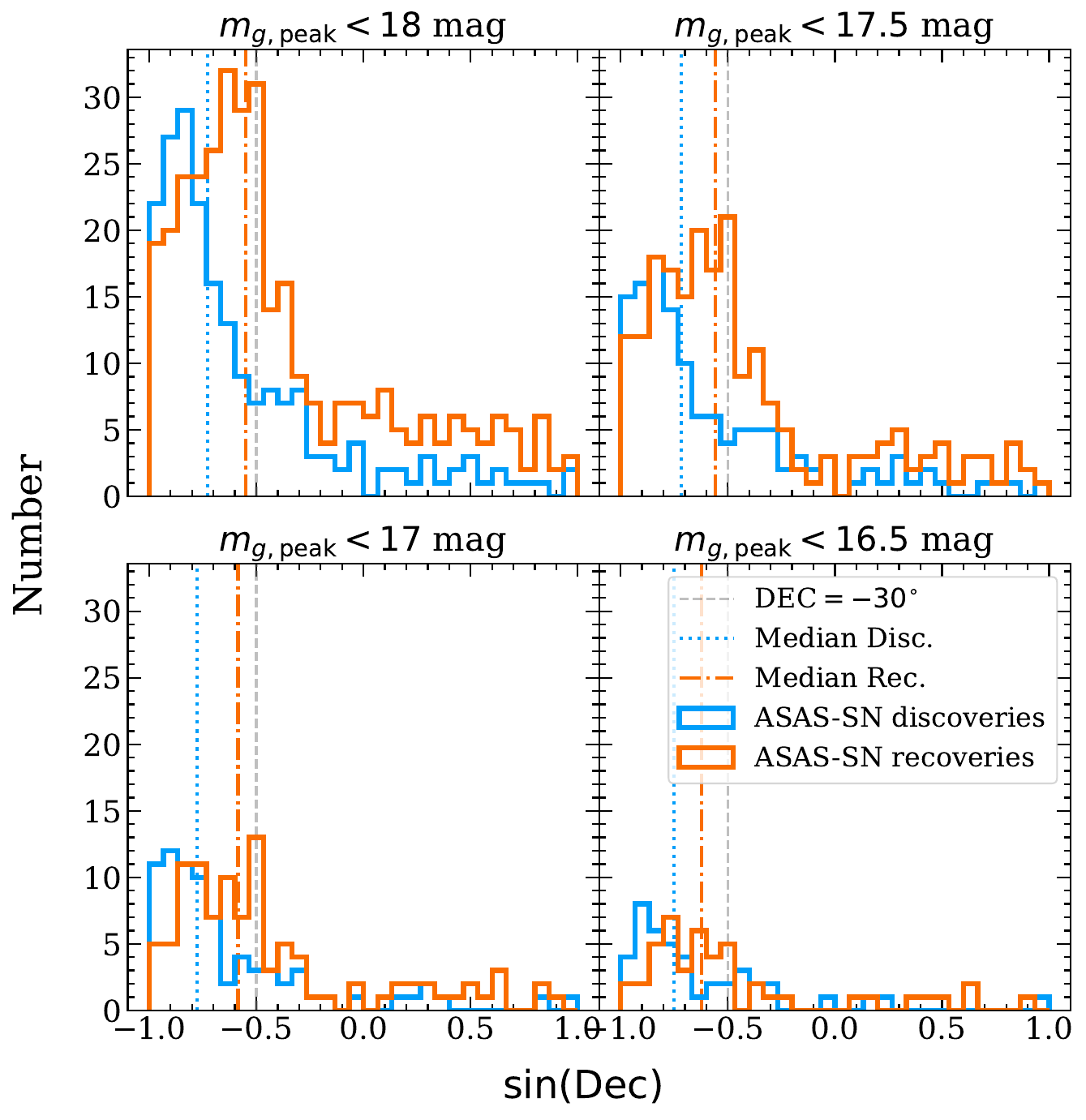}
    \caption{Declination distribution of unclassified transient candidates in the ASAS-SN $g$-band sample for peak magnitude cuts $m_g < 18.0, 17.5, 17.0, 16.5$. The histograms are binned linearly in $\sin \delta$, corresponding to bins of constant sky area. Candidates discovered by ASAS-SN are in blue and those discovered by other surveys but recovered by ASAS-SN are in orange. Vertical dotted lines indicate the median declination for each subsample. The vertical dashed gray line marks $\delta = -30^\circ$, roughly the southern limit for many major Northern hemisphere follow-up facilities. The significant skew toward negative declinations highlights the lower spectroscopic completeness in the Southern hemisphere.}
    \label{fig:unclassified_SNe_dec_dist}
\end{figure}

The spectroscopic incompleteness is not distributed randomly across the sky. As seen in Figure~\ref{fig:unclassified_SNe_dec_dist} the incompleteness is heavily skewed toward the Southern hemisphere ($\delta < 0^\circ$). This strong declination dependence arises from the concentration of follow-up resources in the Northern hemisphere (e.g., SCAT, \citealt{Tucker22a}; the ZTF Bright Transient Survey, \citealt{Fremling20, perley20}). While major classification efforts such as ePESSTO+ \citep{Smartt15} and SCAT South \citep[only since 2023;][]{Tucker22a} target the Southern sky, their combined coverage remains significantly lower than the resources available in the North.

We apply this correction to our rate calculation by introducing the weight $F_{3,i}$ in Equation~\ref{eq:rate}. This assumes that the intrinsic subtype distribution of unclassified SNe within a specific magnitude bin is the same as that of classified SNe. The strong declination dependence of our incompleteness supports the validity of this assumption: the selection function is dominated by telescope visibility limits rather than the physical properties of the supernovae. Therefore, the unobserved sample in the South is likely a random draw from the underlying spectroscopic population, avoiding any biases associated with targeting decisions based on light curve morphology or host galaxy properties.

While our magnitude-dependent correction statistically accounts for these candidates in the corresponding magnitude bins, this disparity underscores the need for increased spectroscopic follow-up resources in the Southern hemisphere to reduce the need for such statistical corrections in future surveys.

\section{Conversion to \texorpdfstring{\lowercase{$g$}-band}{g-band}} \label{app:g_band_convert}
To integrate the $V$-band sample from our earlier ASAS-SN analyses with our current $g$-band sample, and to compare our results with the $r$-band LF from \citet{perley20}, we require a method to convert peak magnitudes between these filters. Since ASAS-SN obtains single-band light curves, we cannot derive colors directly from our data. Furthermore, because SNe~Ia peak at different epochs in different filters but LFs are done in terms of peak magnitude, we calculate the difference in peak magnitudes (e.g., $m_{g,\rm peak} - m_{V,\rm peak}$) rather than a standard color. We derived empirical transformations for these quantities using multiband photometry from CSP~DR3 \citep[][]{Krisciunas17}.

We constructed a set of 76 SNe~Ia from CSP DR3 with well-sampled $g$-, $r$-, and $V$-band light curves. For each SN, we determined the peak apparent and absolute magnitudes in each filter by fitting the light curves with Gaussian Processes. To ensure consistency with our ASAS-SN sample, we corrected the CSP photometry for Milky Way extinction and $K$-corrections, but we did not apply corrections for host-galaxy extinction. This approach preserves the intrinsic scatter due to host reddening that is present in our ASAS-SN dataset which also does not correct for host reddening.

\begin{figure}
    \centering
    \includegraphics[width=\linewidth]{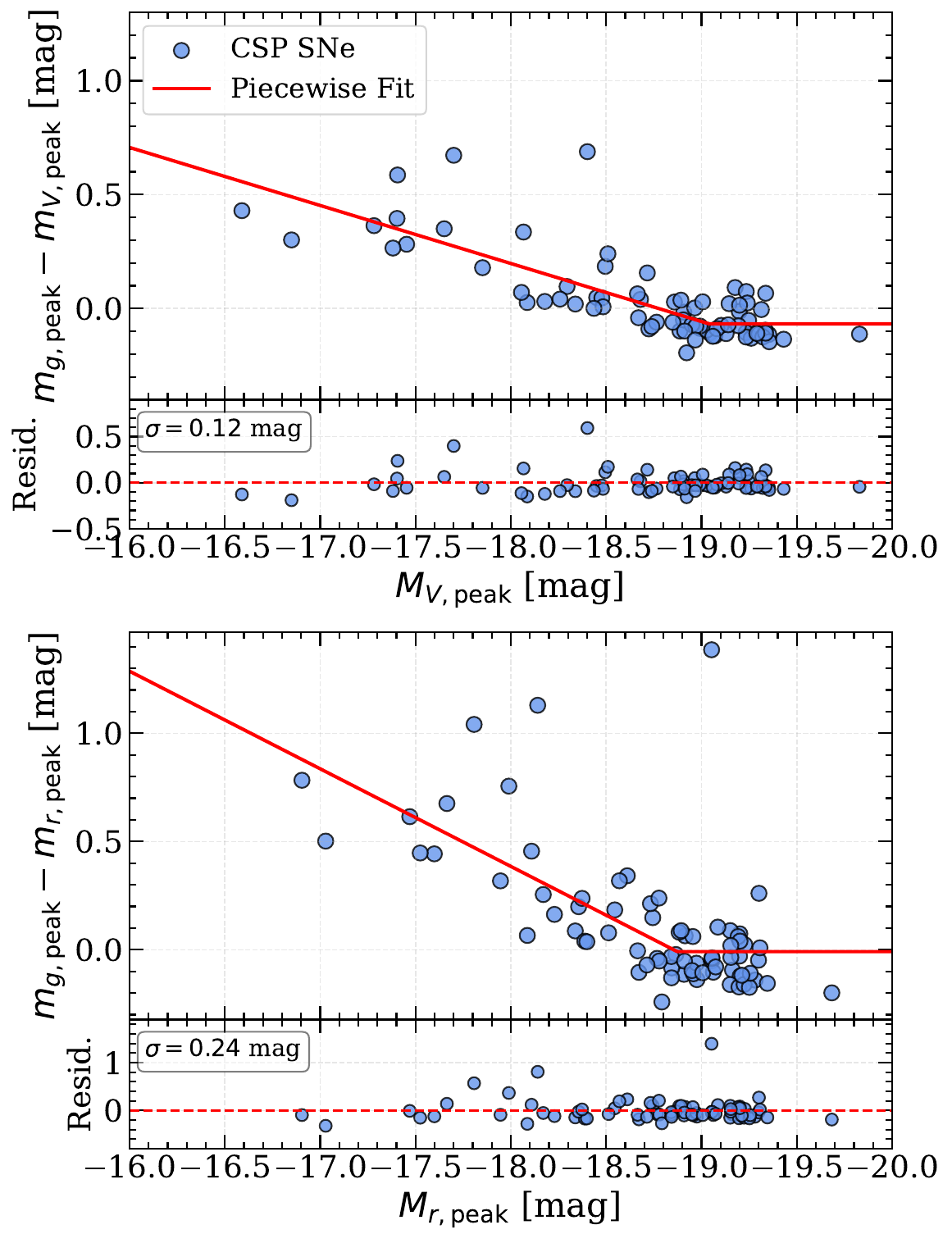}
    \caption{Empirical magnitude transformations derived from the CSP DR3 sample \citep{Krisciunas17}. The blue points represent individual SNe~Ia from the CSP DR3 dataset, processed with Gaussian Process fits and corrected for Milky Way extinction and $K$-corrections (but not host extinction). The solid red lines indicate the piecewise linear fits used to convert $V$ and $r$-band peak magnitudes to the $g$-band.}
    \label{fig:g_band_convert}
\end{figure}

Figure~\ref{fig:g_band_convert} shows the relationship of the peak magnitude differences ($m_{g,\rm peak} - m_{V,\rm peak}$ and $m_{g,\rm peak} - m_{r,\rm peak}$) against the peak absolute magnitudes ($M_{V, \rm peak}$ and $M_{r, \rm peak}$). Bright, standardizable SNe~Ia exhibit a relatively constant difference between peak magnitudes, while lower-luminosity events show a trend toward larger, redder differences.

To model this behavior, we fit a simple piecewise linear function to the data. The model is constrained to be constant for luminous SNe and increases linearly towards lower luminosities. The resulting relations for the $V$ and $r$ bands are given by
\begin{equation}
    \Delta m_{gV} = 
    \begin{cases} 
        -0.067, & M_{V} < -19.042 \\
        -0.067 + 0.255(M_{V} + 19.042), & M_{V} \geq -19.042
    \end{cases}
    \label{eq:g_V_transform}
\end{equation}
\noindent and
\begin{equation}
    \Delta m_{gr} = 
    \begin{cases} 
        -0.009, & M_{r} < -18.873 \\
        -0.009 + 0.451(M_{r} + 18.873), & M_{r} \geq -18.873
    \end{cases}
    \label{eq:g_r_transform}
\end{equation}
\noindent where $\Delta m_{gV} = m_{g,\rm peak} - m_{V,\rm peak}$, $\Delta m_{gr} = m_{g,\rm peak} - m_{r,\rm peak}$, and $M_V$ and $M_r$ are the peak absolute magnitudes in their respective filters.

These relations, shown as the red lines in Figure~\ref{fig:g_band_convert}, serve as the conversion functions for transforming $V$-band and $r$-band peak absolute magnitudes into the $g$-band baseline used in this work. The scatter from the residuals is 0.12~mag for $\Delta m_{gV}$ and 0.24~mag for $\Delta m_{gr}$.

\end{document}